\documentclass[aps,preprint,showpacs,superscriptaddress,groupedaddress]{revtex4}  % for double-spaced preprint
\usepackage{graphicx}  % needed for figures
\usepackage{dcolumn}   % needed for some tables
\usepackage{bm}        % for math
\usepackage{amssymb}   % for math
\usepackage{amsmath} 
\usepackage{gensymb}
\usepackage{vector}

\newcommand{\tensG}{\bvec{\bar{ \bar{G}}}}
\newcommand{\tensI}{\bvec{\bar{ \bar{I}}}}
\newcommand{\tenseps}{\bar{\bar{\boldsymbol{\epsilon}}}}
\newcommand{\tensepsref}{\bar{\bar{\boldsymbol{\epsilon}}}_{ref}}
\newcommand{\tensxi}{\bar{\bar{\boldsymbol{\xi}}}}
\newcommand{\opOmega}{\bar{\bar{\boldsymbol{\Omega}}}}
\newcommand{\Fvec}{\bvec{F}}
\newcommand{\Avec}{\bvec{A}}
\newcommand{\Evec}{\bvec{E}}
\newcommand{\Cvec}{\bvec{\Gamma}}
\newcommand{\Bvec}{\bvec{B}}
\newcommand{\Hvec}{\bvec{H}}
\newcommand{\Dvec}{\bvec{D}}
\newcommand{\rhovec}{\boldsymbol{\rho}}
\newcommand{\veckappa}{\boldsymbol{\kappa}}
\begin{document}

% The following information is for internal review, please remove them for submission
\widetext
%\leftline{Version of \today}
%\leftline{Primary authors: Joe E. Physics}
%\leftline{To be submitted to PRB}
%\leftline{Comment to {\tt d0-run2eb-nnn@fnal.gov} by xxx, yyy}
%\centerline{\em D\O\ INTERNAL DOCUMENT -- NOT FOR PUBLIC DISTRIBUTION}

% the following line is for submission, including submission to the arXiv!!
%\hspace{5.2in} \mbox{Fermilab-Pub-04/xxx-E}

\title{Approached vectorial model for Fano resonances in guided mode resonance gratings}
%\input author_list.tex       % D0 authors (remove the first 3 lines
                             % of this file prior to submission, they
                             % contain a time stamp for the authorlist)
                             % (includes institutions and visitors)
\date{\today}

\author{A.-L. Fehrembach*}\affiliation{Aix-Marseille Univ, CNRS, Centrale Marseille, Institut Fresnel, Marseille, France}
\author{B. Gralak}\affiliation{Aix-Marseille Univ, CNRS, Centrale Marseille, Institut Fresnel, Marseille, France}
\author{A. Sentenac}\affiliation{Aix-Marseille Univ, CNRS, Centrale Marseille, Institut Fresnel, Marseille, France}

\begin{abstract}
We propose a   self-consistent vectorial method, based on a Green's function technique, to describe the Fano resonances that appear in guided mode resonance gratings. The model provides  intuitive expressions of the reflectivity and transmittivity matrices of the structure, involving coupling integrals between the modes of a planar reference structure and radiative modes. These expressions are used to derive a physical analysis in configurations where the effect of the incident polarization is not trivial. We provide numerical validations of our model.    
On a technical point of view, we show how the Green's tensor of our planar reference structure can be expressed as two scalar Green's functions, and how to deal with the singularity of the Green's tensor.

\vspace{1cm}

\noindent\textit{*Corresponding author: anne-laure.fehrembach@fresnel.fr}
\end{abstract}

\maketitle

\section{Introduction}

Fano resonances occur when a discrete resonance state is superposed with broadband continuum states. They are characterized by an asymmetric line shape which is due to the successive destructive and constructive interferences arising between the two generated waves. In the field of photonics, a growing interest in Fano resonances  has been observed this last decade 
with the revelation of their potential for applications in filtering, chemical and biological sensing, light handling, harvesting or absorption \cite{Khanikaev_Nanophot_2013}. Recently, Fano resonances have been generated with plasmonic nanostructures and metamaterials \cite{Lukyanchuk_NatureMat_2010,Rahmani_LaserPhot_2013}, and in structures containing graphene \cite{Gande_OptExpr_2015,Liu_OptExpr_2015}.

In fact, the first reported asymmetric line shape spectrum was observed fortuitously by Wood with light on metallic gratings in 1902 \cite{Wood_1902}. At that time, Wood noticed the physical interest of this unexplained phenomenon termed 'anomaly', but the astronomers avoided it because it hindered their observations. The  fact that the excitation of eigen modes of the structure plays a role in  'Wood's anomalies' was first suggested by  Fano  in 1941 \cite{Fano_1941}, and modelized by Hessel and Oliner in 1965 \cite{Hessel_1965}. On metallic gratings, the eigen modes are surface plasmons. The same phenomenon can be observed with all dielectric structures supporting guided modes \cite{Neviere_OptComm_1973,Peng_IEEE_1975}. The main interest of dielectric gratings with respect to their metallic counterpart is  the possibility to use lossless materials.  In that case, the resonance can be very narrow and remarkably, the reflectivity and transmittivity can reach 100\% provided  that the structure satisfy appropriate symmetry conditions \cite{Popov_optacta_1986}. 
For these reasons, guided mode resonance gratings are particularly attractive for filtering applications \cite{Golubenko_OQE_1986,Wang_ApplOpt_1993}.

However the way the Fano resonance is generated, depicting it with a simple model is of great interest, as an intuitive understanding facilitates the control of its properties (bandwidth, position and amplitude). In this respect, the quasi-normal mode approach developed recently \cite{Sauvan_PRL_2013,Bai_OptExpr_2013,Sauvan_PRA_2014,Vial_PRA_2014,Grigoriev_PRA_2013}  brings a useful  physical insight into the properties of resonant structures. In particular, these properties can be expressed in terms of coupling integrals involving the modes.  
Additional physical insight can be brought by perturbation theories, where the studied structure is seen as a reference structure whose eigen modes are modified by a perturbation, leading to simple expressions of the modification of its properties  \cite{Yang_NanoLett_2015}.
This approach is particularly suitable for guided modes resonance gratings, the reference structure being a planar waveguide (with normal modes),  and the perturbation the grating. It can be based either on the Coupled Mode Method \cite{Norton_JOSAA_1997,Blanchard_PRA_2014}, very popular in the field of integrated optics, or on the Green's function formalism \cite{Evenor_EurPhysJD_2012,Paddon_OptLett_1998}.  The approached modelling of guided mode resonance gratings has been helpful in the design of complex structures, such as the 'bi-atomic grating', a solution to enhance the angular tolerance of the resonance \cite{Lemarchand_OptLett_1998,Sakat_OptLett_2013}. 

Yet, most of the perturbative methods developed for guided mode resonance gratings treat the scalar problem (typically, a 1D grating illuminated along a direction of periodicity). Now, the vectorial case (1D grating illuminated under conical incidence, or 2D grating) presents a strong interest, especially when polarization independent configurations are sought \cite{Mizutani_JOSAA_2001,Fehrembach_JOSAA_2003,Lacour_JOSAA_2003, Niederer_OptExpr_2005,Fehrembach_OptLett_2011}. The complexity of the behavior of resonant structures with respect to the incident polarization is then a strong incentive for developing an approximate model giving a physical insight of the vectorial resonance phenomenon \cite{Alaridhee_OptExpr_2015,Fehrembach_JOSAA_2017}. 

As demonstrated in \cite{Fehrembach_JOSAA_2002}, an efficient tool to study the behavior with respect to the incident polarization of any specularly diffracting structure is the set of eigenvalues of the  reflectivity and transmittivity matrices in energy: they are the  bounds of the reflectivity and transmittivity when the incident polarization takes any elliptical state. The associated eigenvectors correspond to the polarizations for which these bounds are reached. This approach is particularly powerful when the involved modes have non-trivial polarizations \cite{Alaridhee_OptExpr_2015,Fehrembach_JOSAA_2017}. Yet, 
because the involved matrices contain both the resonant and the non-resonant parts of the diffracted field, the relation between  the  excited mode   and the resonance of the eigenvalues is not intuitive. In this paper, we develop a vectorial  approached model for guided mode resonance gratings to fill up this lack of physical insight. 

A vectorial  model based on the Green's tensor formalism was presented in  \cite{Paddon_PRB_2000}, but for the homogeneous problem and not the diffraction problem.  The method we propose here is also based on the Green's tensor formalism and can be seen as further developments of \cite{Evenor_EurPhysJD_2012} (which solves the scalar diffraction problem) and  \cite{Paddon_PRB_2000} (which solves the vectorial homogeneous problem). 

In the first section, we present the Green's tensor formalism to obtain a rigorous integral equation of the diffraction problem. Then, we introduce the guided modes of a  reference planar structure by expanding the Green's tensor on its eigen modes. Making suitable assumptions, we obtain simplified expressions for the reflectivity and transmittivity matrices of the guided mode resonance grating, involving coupling integrals between the guided modes and the radiative modes. In the third section, we derive  physical interpretations from these formulas and, in the fourth section, we provide a numerical validation of our model.

%%%%%%%%%%%%%%%%% SECTION 1
\section{Rigorous integral equation}

\subsection{Geometry of the problem and notations}
\begin{figure}
\includegraphics[width=8cm]{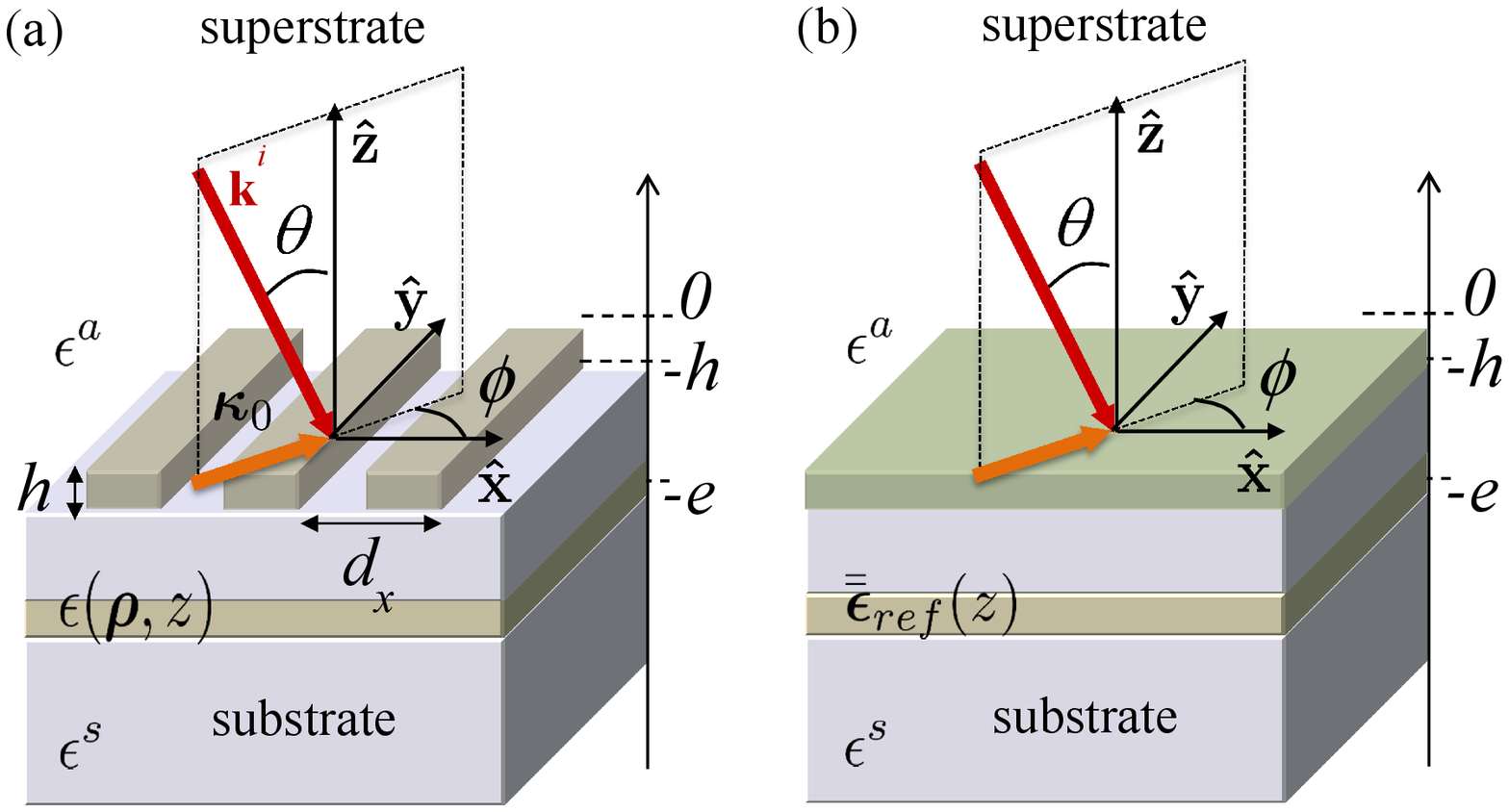}
\caption{(a) Studied configuration and notations. (b) Reference planar structure.}
\label{fig configuration}
\end{figure}

A cartesian coordinate system $(x,y,z)$ is used, the unit vectors being $\buvec{x}$, $\buvec{y}$ and $\buvec{z}$. 
As shown in Fig. \ref{fig configuration}(a), the structure is composed with a stack of homogeneous layers of dielectric materials, which are supposed lossless and infinite along the $x$ and $y$ directions.
A grating is engraved on top of the stack. The grating can be either periodic along the $x$ direction only (1D grating), or along both $x$ and $y$ directions (2D grating). The period along $x$ and potentially $y$ are denoted $d_x$ and $d_y$ respectively. The grating pattern is composed with holes, the shape of which is invariant along the $z$ direction between the planes defined by $z=-h$ and $z=0$.
The relative permittivity of the studied structure is noted $\epsilon(\rhovec,z)$ with $\rhovec=x\buvec{x}+y\buvec{y}$. It is equal to $\epsilon^a$ in the superstrate ($z>0$) and $\epsilon^s$ in the substrate ($z<-e$), where $e$ is the total thickness of the stack, including the engraved layer and without the substrate and the superstrate, which are semi-infinite. 

Throughout the paper, we consider harmonic fields with pulsation $\omega$ and wavelength in vacuum denoted $\lambda$,  with temporal dependency $\exp{(-i\omega t)}$. The structure is illuminated with an incident plane wave coming from the superstrate. $\theta$ is the polar angle of incidence with respect to $\buvec{z}$, and $\phi$ the azimuthal angle of incidence with respect to $\buvec{x}$ (see Fig. \ref{fig configuration}(a)). The projection of the incident wavevector $\bvec{k}^i$ on the $(x,y)$ plane is denoted $\veckappa_0$.  The projection, on the $(x,y)$ plane, of the wavevector of the $m^{\text{th}}$ order diffracted wave is
\begin{equation}
\veckappa_m=\veckappa_0+\bvec{K}_m,
\end{equation}
where $\bvec{K}_m$ is the vector of the reciprocal space of the grating associated with the $m^{\text{th}}$ diffraction order. More precisely, for a grating periodic along the $x$ direction only, $\bvec{K}_m = m \frac{ 2 \pi}{d_x}\buvec{x}$. For a grating periodic along both $x$ and $y$ directions, the integer $m^{\text{th}}$ is associated with the couple of the two relative integers labelling the $(m_x,m_y)$ diffraction order and  $\bvec{K}_m= m_x\frac{ 2 \pi}{d_x}\buvec{x}+ m_y\frac{ 2 \pi}{d_y}\buvec{y}$. The zero diffraction order corresponds to  $m=0$. Throughout the paper, we will consider configurations where the zero order is the only propagative order in the substrate and the superstrate. 

For the sake of clarity, we consider structures containing a single grating, on top of the stack, and isotropic materials only. But the method can be easily extended to structures containing several gratings, whatever their location inside the stack provided that the whole structure is still periodic, and also containing homogeneous anisotopic layers, with $z$ symmetry axis.

\subsection{Set of differential coupled equations}
The electric field $\Evec$ is the solution of the equation
\begin{equation}
\label{Helmholtz}
\nabla \times \nabla \times {\Evec}-k_0^2\epsilon \Evec = \bvec{0},
\end{equation}
where $k_0=2\pi/\lambda$ is the wavenumber in vacuum. Since the structure is periodic along $x$ and possibly $y$, the electric field is pseudo-periodic and can be written as a Floquet-Bloch expansion with coefficients $\Evec_m(z)$
\begin{equation}
\label{Floquet_Bloch}
\Evec(\bvec{\rhovec},z)=\sum_m \Evec_m(z) \exp{(i\bvec{\veckappa_m}.\bvec{\rhovec})}.
%\Bvec(\bvec{\rhovec},z)=\sum_m \Bvec_m(z) \exp{(i\bvec{\veckappa_m}.\bvec{\rhovec})}& 
\end{equation}
We also expand the relative permittivity of the studied structure as a Fourier series, with coefficients $\epsilon_m(z)$:
\begin{equation}
\epsilon(\bvec{\rhovec},z)=\sum_m \epsilon_m(z) \exp{(i\bvec{K_m}.\bvec{\rhovec})}.
\label{Fourier epsilon}
\end{equation}
Inserting eqs. (\ref{Floquet_Bloch}) and (\ref{Fourier epsilon}) into eq. (\ref{Helmholtz}) leads to a set of differential equations coupling the diffraction orders:
\begin{equation}
  \opOmega_m[\Evec_m(z)]-k_0^2  \sum_n \epsilon_{m-n}(z) \Evec_n(z)=\bvec{0}, 
  \label{Helmholtz en z}
\end{equation}
where the operator $\opOmega_m$ is given by
\begin{equation}
  \opOmega_m[\Evec_m(z)]=i\bvec{\veckappa_m}\times(i\bvec{\veckappa_m}\times\Evec_m)+ \buvec{z} \times (\buvec{z} \times \partial_z^2\Evec_m) + \\ (i\veckappa_m.\partial_z\Evec_m) \buvec{z} + (\partial_z\Evec_m . \buvec{z}) i\veckappa_m.
  \label{opOmega}
  \end{equation}
  In the following paragraph, we derive an integral formulation from this equation, introducing the Green's tensor for a reference planar structure.

\subsection{The reference problem}
We consider a structure, called "reference structure", composed with the same homogeneous layers as the studied structure (see Fig. \ref{fig configuration}(b)). In the grating region, to form the reference structure, the grating is replaced with an homogeneous layer with the same thickness, made of a material which can be anisotropic with symmetry axis $z$. This anisotropy yields a supplementary degree of freedom in the model without making too complex the calculations.  The relative permittivity of the reference structure is denoted  $\tenseps^{ref}$. We note $\epsilon^o$ the permittivitty in the $(x,y)$ plane and $\epsilon^e$ along the $z$ axis:
\begin{equation}
\tenseps^{ref}(z)=\epsilon^o(z) (\buvec{x}\buvec{x} + \buvec{y}\buvec{y}) + \epsilon^e(z) \buvec{z}\buvec{z}.
\label{tenseur eps}
\end{equation}
Outside the grating region, $\epsilon^o(z)=\epsilon^e(z)=\epsilon(z)$, as the homogeneous  materials of the studied structure are supposed isotropic.
We will specify the expression of $\tenseps^{ref}$, depending on  the relative permittivity of the grating, in the paragraph II.F.

We consider that the reference structure is illuminated with the same plane wave as the studied structure, with in-plane wavevector $\veckappa_0$ and wavelength $\lambda$. Hence, the field solution of the diffraction problem is  $\Evec^{ref}_0(z) \exp{(i\veckappa_0.\bvec{\rhovec})}$ where $\Evec^{ref}_0(z)$ is the solution of  the following equation for $m=0$:
\begin{equation}
  \opOmega_m[\Evec^{ref}_m(z)]-k_0^2  \tenseps^{ref}(z) \Evec^{ref}_m(z)=\bvec{0}.
  \label{Helmholtz reference}
  \end{equation}
On the other hand, a guided mode of the reference structure is expressed as $\Avec_m(z)\exp{(i \veckappa_m.\rhovec)}$, where $\Avec_m(z)$ is solution of the homogeneous eq. (\ref{Helmholtz reference}) for $m \neq 0$ since the zero diffraction order is propagating in the substrate and superstrate, while the guided modes must be evanescent in those media.

We now introduce the Green's tensor $\tensG_m(z,z')$, associated to the reference structure, as the solution of 
\begin{equation}
  \opOmega_m[\tensG_m(z,z')]-k_0^2\tenseps^{ref}(z) \tensG_m(z,z')=k_0^2  \delta(z-z') \tensI, 
  \label{Helmholtz Green}
  \end{equation}
and satisfying the outgoing wave condition. $\tensI$ is the identity tensor, and $\delta(z-z')$ is the Dirac distribution. The index $m$ indicates that the considered in-plane wavevector is $\veckappa_m$. 

As the reference structure is anisotropic with symmetry axis $z$, eqs. (\ref{Helmholtz reference}) and  (\ref{Helmholtz Green}) can be expressed as two scalar problems for the two fundamental polarizations, transverse electric field (TE) and transverse magnetic field (TM) with respect to the direction of propagation, as shown in appendix A and B. 

\subsection{Rigorous integral equation}
Once the Green's tensor solution of eq. (\ref{Helmholtz Green}) is known, the differential eq. (\ref{Helmholtz en z}) can be transformed into an integral equation. 
Eq. (\ref{Helmholtz en z}) can be written as 
\begin{equation}
  \opOmega_m(\Evec_m)-k_0^2 \tenseps^{ref}\Evec_m  = k_0^2 \sum_n \left[ \epsilon_{m-n}\tensI -\delta_{m,n}\tenseps^{ref}\right]\Evec_n,
  \label{Helmholtz en z 2}
\end{equation}
where $\delta_{m,n}$ is the Kronecker symbol.
By subtracting eq. (\ref{Helmholtz reference}) to eq. (\ref{Helmholtz en z 2}), we obtain
\begin{equation}
  \opOmega_m(\Evec^{pert}_m)-k_0^2 \tenseps^{ref}\Evec^{pert}_m  = k_0^2 \sum_n \left[ \epsilon_{m-n}\tensI -\delta_{m,n}\tenseps^{ref}\right]\Evec_n, 
  \label{Helmholtz en z 3}
\end{equation}
where  we introduced the field  $\Evec^{pert}_m$ defined by
\begin{equation}
\Evec^{pert}_m=\Evec_m-\Evec^{ref}_m.
\label{Epert}
\end{equation}
 %$\delta_{m,0}$ being the  Kronecker symbol.
  It has to be noted that, since $\Evec_m$ and $\Evec^{ref}_m$ are generated by the same incident field, $\Evec^{pert}_m$ satisfies the outgoing wave condition. 
From eq. (\ref{Helmholtz en z 3}) we deduce, using eq. (\ref{Helmholtz Green}) that 
     \begin{equation}
  %\Evec_m(z)=\Evec^{ref}_0(z)\delta_{m,0}+ \int_{-h}^{0} \text{d}z'   \tensG_m(z,z') \sum_n (\epsilon_{m-n}(z')\tensI-\delta_{m,n} \tensepsref(z'))\Evec_n(z'),%
    \Evec_m(z)=\Evec^{ref}_m(z)+ \int_{-h}^{0} \text{d}z'   \tensG_m(z,z') \sum_n \left[\epsilon_{m-n}(z')\tensI-\delta_{m,n} \tensepsref(z')\right]\Evec_n(z'),%
  \label{integrale Etot}
  \end{equation}
  where we used that the reference structure and the studied structure have the same permittivity outside the grating region (the grating region is for $z\in[-h,0]$).
  The calculation of the Green's tensor detailed in the appendix B shows that it presents a singularity on its $\buvec{z}\buvec{z}$ component (the same singularity as for layered isotropic materials \cite{Gralak_JMathPhys_2010}), and can be written as
\begin{equation}
\tensG_m(z,z')=\tensG^{NS}_m(z,z')-\frac{1}{\epsilon^e} \delta(z-z') \buvec{z}\buvec{z},
\label{G sing}
\end{equation}
where $\tensG^{NS}_m$ is the non-singular part of $\tensG_m$. We show in the following paragraph how we can deal with this singularity.
 
\subsection{Treatment of the singularity of the Green's tensor}
The calculation of the integral over $z$ of the singularity in eq. (\ref{integrale Etot}) leads to the term 
\begin{equation}
-\frac{\buvec{z}\buvec{z}}{\epsilon^e}\sum_n (\epsilon_{m-n}(z)\tensI-\delta_{m,n} \tensepsref(z))\Evec_n(z).
\label{terme sing}
\end{equation}
By transferring this term to the left hand side of eq. (\ref{integrale Etot}) we can write 
\begin{equation}
  \Fvec_m(z)=\Evec^{ref}_m(z)  +  \int_{-h}^{0} \text{d}z'   \tensG^{NS}_m(z,z') \sum_n (\epsilon_{m-n}(z')\tensI-\delta_{m,n} \tensepsref(z'))\Evec_n(z'),
  \label{aux1}
  \end{equation}
where
\begin{equation}
\displaystyle{\Fvec_m(z)=\sum_n \left[\delta_{m,n}(\buvec{x}\buvec{x}+\buvec{y}\buvec{y})+\frac{\epsilon_{m-n}(z)}{\epsilon^e(z)}\buvec{z}\buvec{z}\right]\Evec_n(z)}.
\label{expr F}
 \end{equation}
From eq. (\ref{expr F}), we can express $\Evec_n$ as a function of $\Fvec_m$:
\begin{equation}
\displaystyle{\Evec_n(z)=\sum_m \left[\delta_{m,n}(\buvec{x}\buvec{x}+\buvec{y}\buvec{y})+\epsilon^e(z)\left[\frac{1}{\epsilon(z)}\right]_{n-m}\buvec{z}\buvec{z}\right]\Fvec_m(z)},
 \end{equation}
where $\left[\frac{1}{\epsilon(z)}\right]_{p}$ is the $p^{\text{th}}$ coefficient of the Fourier expansion of the function $1/\epsilon(z)$.
Last, by replacing $\Evec_n$ by this expression into eq. (\ref{aux1}), we obtain
\begin{equation}
  \Fvec_m(z)=\Evec^{ref}_m(z)  +  \int_{-h}^{0} \text{d}z' \tensG^{NS}_m(z,z') \sum_n \tensxi_{m-n}(z')\Fvec_n(z'),
  \label{eq integrale F}
  \end{equation}
where $\tensxi_{m-n}(z)$ is defined by
 \begin{equation}
 \tensxi_{m-n}(z)=(\epsilon_{m-n}(z)-\epsilon^o(z)\delta_{m,n})(\buvec{x}\buvec{x}+\buvec{y}\buvec{y})+\epsilon^e(z)\left(\delta_{m,n}-\epsilon^e(z)\left[\frac{1}{\epsilon(z)}\right]_{m-n}\right)(\buvec{z}\buvec{z}),
 \label{eq def xi}
 \end{equation}
 taking into account that $\sum_p\left[\epsilon\right]_{m-p}\left[\frac{1}{\epsilon}\right]_{p-n}=\delta_{m,n}$.
   It is useful to note that   $\Fvec_m$ is equal to $\Evec_m$ outside the grating region. In the grating region, they differ only for the $\buvec{z}$ component.
      Eq. (\ref{eq integrale F}) represents the coupling between the $m^{\text{th}}$ order and the $n^{\text{th}}$ order, through the coefficient $\tensxi_{m-n}$, representing the perturbation induced by the grating on the reference structure.

\subsection{Choice of the reference structure}
We choose the planar reference structure such that it gives a diffracted field as close as possible to the field diffracted by the considered structure.
In other words, the perturbation, represented by the coefficients $\tensxi_{p}$, must not allow the direct coupling from one order to itself. This condition mathematically translates into $\tensxi_{0}=0$. From the expression of $\tensxi_{m-n}$ (eq. (\ref{eq def xi})), we deduce that 
\begin{equation}
\epsilon^o=\epsilon_m \delta_{m,0} \quad \quad \text{and} \quad \quad
\epsilon^e=\left[\frac{1}{\epsilon(z)}\right]_{m}^{-1}\delta_{m,0}
\end{equation}
Hence, the ordinary permittivity $\epsilon^o$ must be equal to the geometric average of the grating permittivity, and the extraordinary permittivity $\epsilon^e$ must be equal to its harmonic  average. First of all, note that this result directly follows from our  hypothesis for the reference structure to be anisotropic with $z$-axis. It also comes from the fact that we first performed a Fourier transform in the $(x,y)$ plane, leading to a perturbative model with respect to the grating depth $h$ (see Eq. \ref{eq integrale F}), and to an homogenization of the grating in the limit of small depths. As a matter of fact, the singularity of our Green tensor is the same as the singularity calculated by Yaghjian \cite{Yaghjian_procIEEE_1980} when integrating the Green tensor of an homogeneous medium in the source region over a thin "pillar box" in the $(x,y)$ plane. Moreover, our result is consistent with the well known rules for the homogenization of a periodic assembly of thin plates \cite{Born_and_Wolf}: the effective permittivity is the geometric average of the permittivities for the directions where the electric field is continuous through the plates interfaces (i.e. directions parallel to the plane of the plates), while it is the  harmonic average for the direction where the electric displacements is continuous (i.e. direction perpendicular to the plane of the plates). Usually, the plates are parallel to each others \cite{Born_and_Wolf}, e.g. the direction of periodicity is perpendicular to the plane of the plates. Our case differs since our plates are juxtaposed in the $(x,y)$ plane, e.g. the directions of periodicity are contained in the plane of the plates. Yet, the same rules apply: the direction of the plates imposes the form of the permittivity tensor, while the directions of periodicity give the directions along which the average is performed. The rule of the geometric average in the plane for a small depth 1D or 2D grating was already mentioned in Ref. \cite{Lalanne_JOSAA_1997}.
Last, one could expect that the z-axis anisotropic reference structure is more suitable to model 2D gratings (with $\pi/2$ rotation invariance around $z$) than 1D gratings, as the latter creates a strong form anisotropy in the $(x,y)$ plane. Yet, it is important to note that the guided modes excited propagate in directions close to the directions of periodicity of the grating (due to the coupling condition). As a consequence, the fact that the 1D grating has a translation invariance along its ridges has a  minor impact on the propagation of the guided mode. Hence, the model has the same accuracy for 1D and 2D gratings, as will be shown by the numerical calculations.

%\textit{POUR ANNE}: je n'ai pas parlé de Maxwell-Garnett comme tu le suggérais, bien que ça donne effectivement le même résultat. A ceci plusieurs raisons: je n'ai pas trouvé de papier à citer où c'est bien fait, je ne pense pas que ça soit une bonne idée de mettre le calcul en annexe, car c'est un peu hors sujet, le papier est déjà long, et en plus, Evgueni est en train de le faire dans un papier en cours sur l'homogénéisation.

%%This result can be analyzed considering the well known rules for the homogenization of 

%A VOIR
%1D periodic structures: the geometric average of the permittivity must be considered in the direction where the electric field is tangential to the interfaces and the harmonic average in the direction where the electric field is orthogonal to the interfaces. For a 1D grating, this leads to a uniaxial anisotropic material with the anisotropy axis along the grating grooves, i.e. with $x$- or $y$- anistropy. A 2D grating would be replaced in general with a biaxial anisotropic material, or, in the case of a unit cell invariant under a $\pi/2$ rotation, by a uniaxial anisotropic material with the $z$-anisotropy axis. It appears that our reference structure does not follow these rules. Yet, as already mentioned, we consider uniaxial anisotropic material with a $z$ anisotropy axis for the sake of simplicity of the calculations. 

It must be noted that eq. (\ref{eq integrale F}) is rigorous. We will now make  approximations in order to obtain an expression of the diffracted field. 

%%%%%%%%%%%%%%%%% SECTION 2
\section{Approached expression of the diffracted field}

We consider that the angles of incidence are fixed, and we are interested in deriving an expression of the diffracted field with respect to the wavelength. We suppose that the reference structure supports guided modes in the range of the considered wavelengths, and that these eigen modes can be excited through diffraction orders of the grating (except the zero order). The coupling condition can be satisfied when the in-plane wavevector of a diffraction order ($\veckappa_q$) has its modulus close to the propagation constant of a guided mode. We note $\mathcal{Q}$ the set of integers $q$ corresponding to the resonant diffraction orders. Depending on the configuration,  $\mathcal{Q}$ can contain only one or several integers. In the following, we treat the general case where $\mathcal{Q}$  contains several integers, the case of a single resonant order being easily deduced from this general case. 

\subsection{Eigen modes of the Green's tensor}

As detailed in the appendix D, based on  results demonstrated in the appendix C, the regular part of the Green's tensor for our planar reference structure can be expanded on the basis of its eigen modes. As a first simplifying hypothesis, we suppose that in the vicinity of the resonance wavelength of one mode, the term corresponding to this mode prevails over the other terms of the sum. Hence, for a resonant order $q \in \mathcal{Q}$, we write, in the vicinity of the resonance wavelength $\lambda_{q}$ of the excited mode,
\begin{equation}
\tensG_q^{NS}(z,z') \simeq \frac{\Avec_{q}(z) \otimes \overline{\Avec_{q}}(z')}{\left[\left(\frac{\lambda}{\lambda_{q}}\right)^2 -1,\right]},
\label{decomp tens G approx}
\end{equation}
where $\otimes$ denotes the tensor product between two vectors, $\Avec_{q}$ is the electric field of the excited guided mode of the reference structure,  $\overline{\Avec_{q}}$ its complex conjugated. This mode is the solution of the homogeneous problem associated with eq.  (\ref{Helmholtz reference}) for $m=q$.
%\begin{equation}
%\tensG_q^{NS}(z,z')=\sum_n\frac{\Avec_{q,n}(z) \otimes \overline{\Avec_{q,n}}(z')}{\left[\left(\frac{\lambda}{\lambda_{q,n}}\right)^2 -1,\right]},
%\label{decomp tens G}
%\end{equation}
%From eq. (\ref{decomp tens G}), we . To simplify the notations, we omit the index $n$ present in the sum of eq. (\ref{decomp tens G}). 
In the following, we will consider that $\lambda$ is always different from any $\lambda_q$. Hence, the only non-null component of the reference field  in eq. (\ref{eq integrale F}) is $\Evec^{ref}_0(z)$.

\subsection{Approached integral equations}
The Green's tensor $\tensG_q^{NS}$ appears as a common factor in the sum contained in the expression of the diffraction order $\Fvec_q$ given by eq. (\ref{eq integrale F}). Injecting eq. (\ref{decomp tens G approx}) into eq. (\ref{eq integrale F}) for $m=q\in \mathcal{Q}$, we obtain
\begin{equation}
  \Fvec_q(z) \simeq \frac{ \Avec_q(z)}{\left[\left(\frac{\lambda}{\lambda_q}\right)^2 -1,\right]} \int_{0}^{-h} \text{d}z' \overline{\Avec_q}(z') . \left[   \sum_{n \notin \mathcal{Q}} \tensxi_{q-n}(z')\Fvec_n(z') + \sum_{q' \in \mathcal{Q}} \tensxi_{q-q'}(z')\Fvec_{q'}(z') \right],
  \label{eq integrale F q}
  \end{equation}
  where we have considered that $q\ne0$ and separated the sums of the resonant and the non-resonant  terms. Note that the tensor product  $\otimes$ is no more necessary in this equation since the quantity under the integral is scalar (scalar product between the vector $\overline{\Avec_q}(z') $ and the vector in the brackets). 
   We can expect that, in the vicinity of $\lambda_q$, the resonant orders $\Fvec_q$ are predominant over the non-resonant orders. Hence, in the sum contained in the expression of a non-resonant order $\Fvec_n$ for $n \notin \mathcal{Q}$, we retain only the resonant orders, as a second simplifying hypothesis:
       \begin{equation}
  \Fvec_n(z) \simeq \Evec^{ref}_0(z)\delta_{n,0}  +  \int_{-h}^{0} \text{d}z'   \tensG^{NS}_n(z,z') \sum_{q\in\mathcal{Q}} \tensxi_{n-q}(z')\Fvec_q(z').
  \label{eq integrale F approx}
  \end{equation}
Once the $\Fvec_q$ are calculated (see the next paragraph), eq. (\ref{eq integrale F approx}) will be used to express the field diffracted in the non-resonant orders, and especially the zero order.

\subsection{Coupling integrals}
The third simplifying hypothesis is to consider that the field in a resonant order $q$ is proportional to the field of the guided mode $\Avec_q$, 
\begin{equation}
\Fvec_q=\sigma_q \Avec_q
\label{proportio}
\end{equation}
with $\sigma_q$ as proportionality coefficient. This hypothesis is suggested by the form of eq.  (\ref{eq integrale F q}), where $\Avec_q(z)$ appears as a factor. One could be tempted to write that $\Fvec_q$ is proportional to $\frac{\Avec_q(z)}{\left[\left(\frac{\lambda}{\lambda_q}\right)^2 -1,\right]}$. Yet, note that the eigen wavelength of the modes of the studied structure are different from that 
of the planar reference structure, hence $\lambda_q$ can not be a pole of $\Fvec_q$. In particular, the  modes of the structure perturbed by the grating are leaky modes. This means that their eigen wavelengths are complex numbers \cite{Neviere_book}. 

Injecting eq. (\ref{proportio}) into eq. (\ref{eq integrale F approx}) leads to
\begin{equation}
  \Fvec_n(z) \simeq \Evec^{ref}_0(z)\delta_{n,0}  +      \sum_{q\in\mathcal{Q}} \sigma_q \Cvec_{n,q}(z),
  \label{eq integrale F approx couplage}
  \end{equation}
  where $\Cvec_{n,q}(z)$  is a vector corresponding to the out coupling out of the mode $q$ through the $n^{\text{th}}$ diffraction order: 
  \begin{equation}
 \Cvec_{n,q}(z) = \int_{-h}^{0} \text{d}z'\tensG^{NS}_n(z,z')\tensxi_{n-q}(z') \Avec_q(z').
 \label{Cvec_nq}
  \end{equation}
  
  In order to calculate the $\sigma_q$ coefficients, we report the expression of $\Fvec_n$ given by  eq. (\ref{eq integrale F approx}) into eq. (\ref{eq integrale F q}), and use the proportionality relation (eq. (\ref{proportio})). We obtain
  \begin{equation}
  \Fvec_q(z) \simeq \frac{ \Avec_q(z)}{\left[\left(\frac{\lambda}{\lambda_q}\right)^2 -1,\right]} \left(  C_{q,0} +  \sum_{q' \in\mathcal{Q}}  \sigma_{q'} \left( C_{q,q'} + \sum_{n \notin \mathcal{Q}}   C_{q,n,q'} \right)\right),
  \label{Fq}
  \end{equation}
where we introduced:
\begin{itemize}
\item{the coupling integral between the reference field and the mode (excitation of the mode):}  
 \begin{equation}
 C_{q,0}=\int_{-h}^{0} \text{d}z' \overline{\Avec_q}(z') \tensxi_{q}(z') \Evec^{ref}_0(z'),
 \label{def Cq0}
 \end{equation}
 \item{the direct coupling integral between the mode $q$ and the mode $q'$:}
  \begin{equation}
 C_{q,q'}=\int_{-h}^{0} \text{d}z' \overline{\Avec_q}(z') \tensxi_{q-q'}(z') \Avec_{q'}(z'),
 \label{def Cqq}
 \end{equation}
 \item{the second order coupling integral between the mode $q$ and the mode $q'$ through the $n^\text{th}$ order:}
 \begin{equation}
 C_{q,n,q'}=\int_{-h}^{0} \text{d}z' \overline{\Avec_q}(z') \tensxi_{q-n}(z') \int_{-h}^{0} \text{d}z'' \tensG^{NS}_n(z',z'')\tensxi_{n-q'}(z'') \Avec_{q'}(z'').
 \label{def Cqnq}
 \end{equation}
 \end{itemize}
  Last, from eq. (\ref{Fq}), we find that the coefficients  $\sigma_q$ are the solutions of the system of linear equations
  \begin{equation}
\sigma_q\left(\left(\frac{\lambda}{\lambda_q}\right)^2 -1-\Sigma_{q,q}\right)- \sum_{q' \in\mathcal{Q}} \sigma_{q'}\left( C_{q,q'} + \Sigma_{q,q'} \right)=C_{q,0}, 
\label{syst lin}
\end{equation} 
where we introduced the notation $\Sigma_{q,q'}=\sum_{n \notin \mathcal{Q}}   C_{q,n,q'}$.
This system of equations shows how the coefficients $\sigma_q$ are modified by the direct coupling between the modes, and also by the second order coupling, which appears in eq. (\ref{syst lin}) as the sum of the integrals $C_{q,n,q'}$. Numerically, we will calculate this sum for $n$ from $-N$ to $N$, and consider $N$ as a convergence parameter. 

\begin{figure}
\includegraphics[width=8cm]{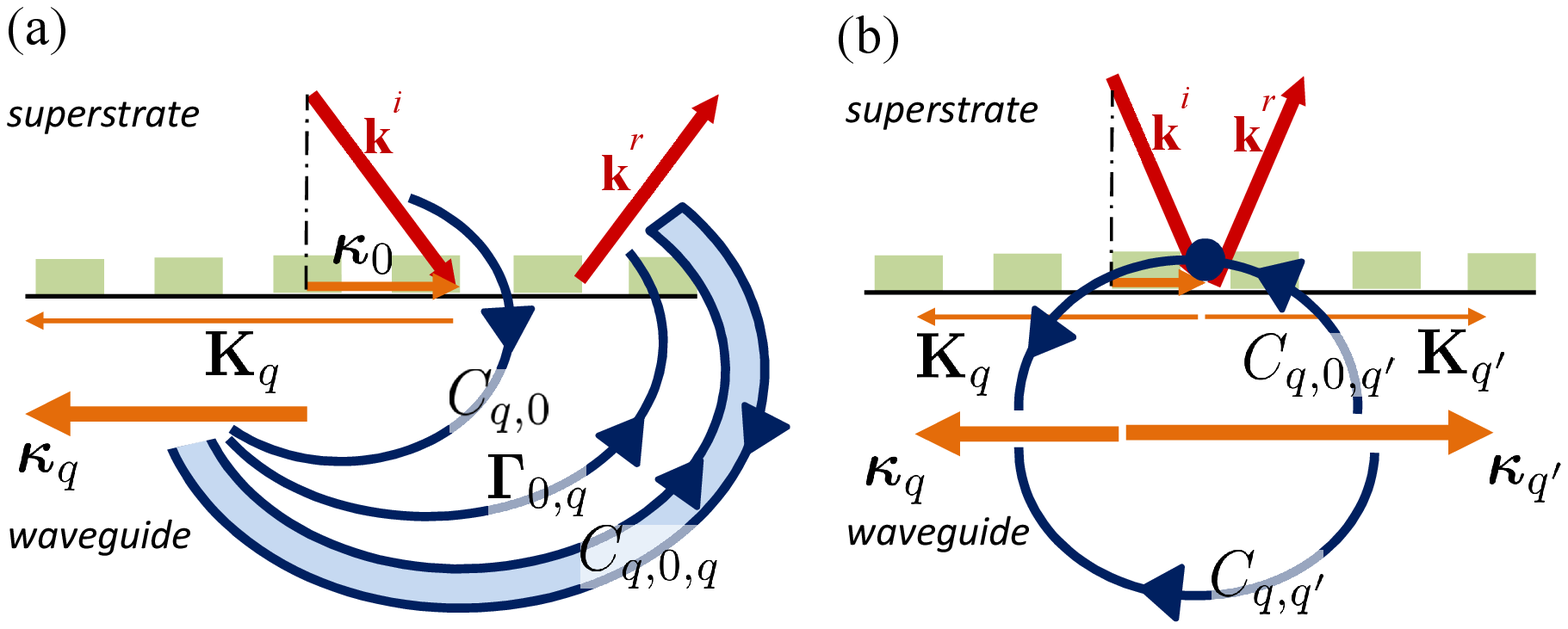} 
\caption{Sketch of the coupling integrals. (a) One resonant order $q$: excitation of the mode ($C_{q,0}$), out-coupling ($\Cvec_{0,q}$) and second order self coupling through the zero order  ($C_{q,0,q}$). (b) Two resonant orders $q$ and $q'$: direct coupling between the two modes ($C_{q,q'}$), and  second order coupling through the zero order ($C_{q,0,q'}$).}
\label{fig couplages}
\end{figure}

We represented in Fig. \ref{fig couplages} a sketch illustrating the couplings, through the coupling integrals (eqs. \ref{Cvec_nq}, and \ref{def Cq0}-\ref{def Cqnq}), in the case of one resonant order only (Fig. \ref{fig couplages}(a)) and two resonant orders (Fig. \ref{fig couplages}(b)).

\subsection{Reflection and transmission matrices}

\begin{figure}[b]
\includegraphics[width=4cm]{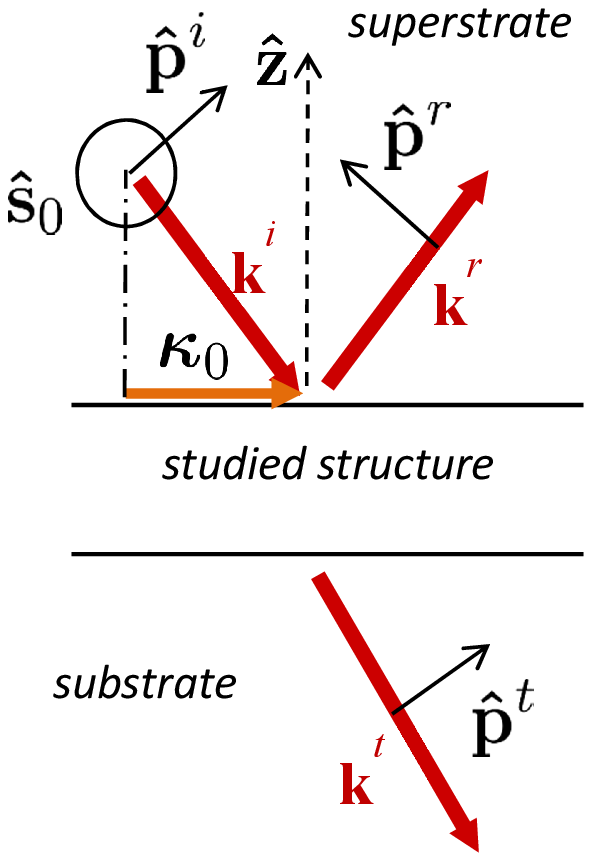}
\caption{$(s,p)$ basis associated with the propagative diffraction orders. }
\label{fig sp basis}
\end{figure}

To express the reflection and transmission matrices of the structure (for the zero diffraction order only), we introduce the basis related to the $s$ and $p$ polarizations. To each diffraction order $m$, we associate the vector $\buvec{s}_m=\buvec{\veckappa}_m \times \buvec{z}$. Moreover, for the incident, reflected and transmitted plane waves with wavevector $\bvec{k}^i$, $\bvec{k}^r$ and $\bvec{k}^t$ respectively, we introduce the vectors $\buvec{p}^i=\buvec{s}_0 \times  \buvec{k}^i$, $\buvec{p}^r=\buvec{s}_0 \times  \buvec{k}^r$ and $\buvec{p}^t=\buvec{s}_0 \times  \buvec{k}^t$ (see Fig.\ref{fig sp basis}). 

The reflection and transmission coefficients can be deduced from eq. (\ref{eq integrale F approx couplage}) for $n=0$. Indeed, since the field  $\Fvec_n$ is equal to $\Evec_n$ outside the grating region, eq. (\ref{eq integrale F approx couplage}) writes,   for $z>0$ and $z<-h$:
\begin{equation}
  \Evec_0(z) \simeq \Evec^{ref}_0(z)\delta_{n,0}  +      \sum_{q\in\mathcal{Q}} \sigma_q \Cvec_{0,q}(z).
  \label{eq integrale E approx couplage hors reseau}
  \end{equation}
For $z>0$, the reference field $\Evec^{ref}_0(z)$ is the sum of the incident field with amplitude $\Evec^{inc}_0$ and the field reflected by the reference structure, with amplitude $\Evec^{ref,r}_0$ :
\begin{equation}
\Evec^{ref}_0(z)=\Evec^{inc}_0\exp{(-i \gamma_a z)}+\Evec^{ref,r}_0\exp{(i \gamma_a z)},
\end{equation}
where $\gamma_a=\sqrt{\epsilon^a k_0^2-\kappa_0^2}$. 
For  $z>0$, the Green's tensor $\tensG^{NS}_0(z,z')$ can be written as $\tensG^{NS}_0(0,z') \exp{(i \gamma_a z)}$, hence
  \begin{equation}
 \Cvec_{0,q}(z) =  \exp{(i \gamma_a z)} \int_{-h}^{0} \text{d}z'\tensG^{NS}_0(0,z')\tensxi_{-q}(z') \Avec_q(z').
  \end{equation}

Taking into account these remarks, we obtain the field  $\Evec^r_0(z)$ (for $z>0$) reflected by the studied structure in the zero order: 
\begin{equation}
\Evec^r_0(z) \simeq \left[\Evec^{ref,r}_0+\sum_{q\in\mathcal{Q}} \sigma_q \Cvec_{0,q}(0)\right]\exp{(i \gamma_a z)}.
\label{eq extration champ inc}
\end{equation}
The vector $\Cvec_{0,q}(0)$ represents  the electric field of the mode in the diffraction order $q$ coupled out by the grating through the zero order. Its components in the ($s,p$) basis are denoted $C^s_{0,q}=\Cvec_{0,q}(0).\buvec{s}_0$ and  $C^p_{0,q}=\Cvec_{0,q}(0).\buvec{p}^r$. 

Now, we consider successively a $s$ and a $p$ polarized incident field, and denote $\sigma^s_q$ and $\sigma^p_q$ the solutions calculated from eq. (\ref{syst lin}) by considering respectively a $s$ and a $p$ incident field in $C_{q,0}$ (eq. (\ref{def Cq0})). We deduce from eq. (\ref{eq extration champ inc}), an approached expression of the reflectivity matrix  $\mathbf{R}$ (zero diffraction order only) of the structure in the ($s,p$) basis
\begin{equation}
\mathbf{R} \simeq \mathbf{R}^{ref}+\sum_{q\in\mathcal{Q}}\mathbf{R}^{pert}_q.
\label{Rref+Rpert}
\end{equation}
The 2x2 matrices $\mathbf{R}$ and $\mathbf{R}^{ref}$  contain the reflectivity coefficients for the studied structure and reference structure respectively, expressed in the  ($s,p$) basis, and the $\mathbf{R}^{pert}_q$ matrices are given by
\begin{equation}
\mathbf{R}^{pert}_q= 
\begin{bmatrix}
\sigma^s_q C^s_{0,q}(0) & \sigma^p_q C^s_{0,q}(0)  \\
\sigma^s_q C^p_{0,q}(0) & \sigma^p_q C^p_{0,q}(0)
\end{bmatrix}.
\label{Rpert}
\end{equation}

Following the same steps for $z<-e$, we obtain an approached expression of the transmittivity matrix  $\mathbf{T}$ (zero diffraction order only) of the structure
\begin{equation}
\mathbf{T}=\mathbf{T^{ref}}+\sum_{q\in\mathcal{Q}} \mathbf{T}^{pert}_q ,
\label{Tref+Tpert}
\end{equation} 
where $\mathbf{T}$ and $\mathbf{T}^{ref}$ are 2x2 matrices containing the transmittivity coefficients for the studied structure and reference structure respectively, expressed in the  ($s,p$) basis and 
\begin{equation}
\mathbf{T}^{pert}_q \simeq 
\begin{bmatrix}
\sigma^s_q C^s_{0,q}(-e) & \sigma^p_q C^s_{0,q}(-e)  \\
\sigma^s_q C^p_{0,q}(-e) & \sigma^p_q C^p_{0,q}(-e)
\end{bmatrix}
\label{Tpert}
\end{equation}
%\sigma_q \Cvec_{0,q}(-e) 
 
 \begin{figure}[b]
\includegraphics[width=8cm]{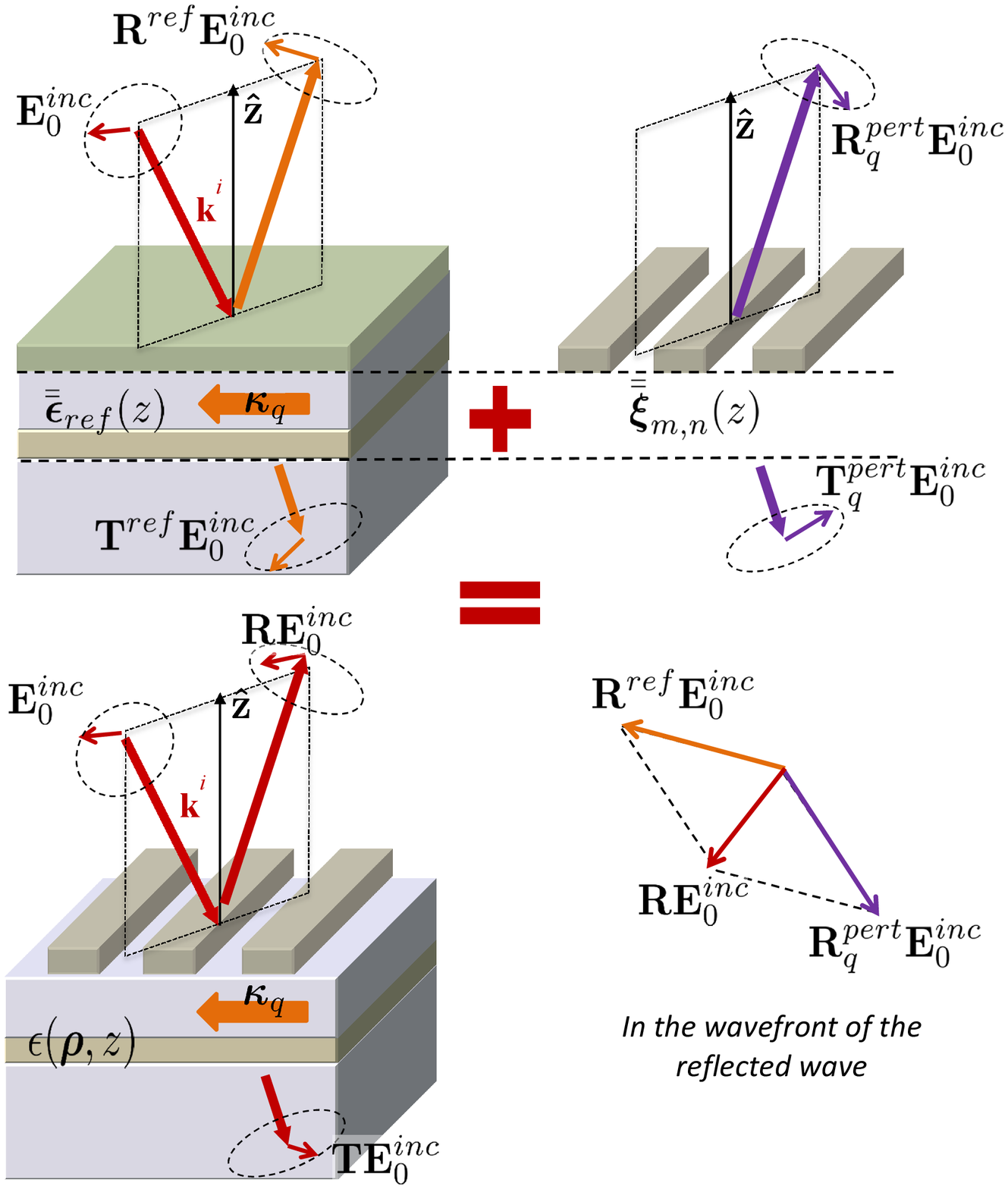}
\caption{At resonance, the reflected and transmitted fields are the sum of a non-resonant term, coming from the planar reference structure, and a resonant term, which is due to the grating. The vectorial sum of the field in the wavefront of the reflected wave is represented on the right bottom sketch.}
\label{fig resume}
\end{figure}
 
The eqs. (\ref{Rref+Rpert}) and (\ref{Tref+Tpert}) appear to be intuitive expressions of the reflectivity and transmittivity matrices of a guided mode resonance grating, where the coupling in and out of a mode appears as an additional term  to the reflectivity and transmittivity matrices of a reference structure. This result is represented by a sketch on Fig. \ref{fig resume}. The addition is expressed in terms of coupling integrals involving the excited modes and radiative modes. The novelty of the present formulation is  that it takes into account the effect of the polarization (of the incident field and of the field diffracted by the studied structure, as well as  that of the mode). We believe in the interest of the vectorial formulation since the behavior of guided mode resonance gratings with respect to the incident polarization may in some configurations  be surprising \cite{Alaridhee_OptExpr_2015,Fehrembach_JOSAA_2017}. 

%The further physical analysis of these formula is outside the scope of this paper. The next section is devoted to their validation.
\section{Physical analysis}

We will consider successively the two situations where one order only is resonant, and then two orders. The former situation corresponds to the general case of oblique incidence. The latter situation will be reduced to the particular case where the plane of incidence is a plane of symmetry of the structure (full conical incidence).  Particular attention will be paid to the influence of the incident polarization.  

\subsection{Situation with one resonant order only}
 
In the situation where only one order is resonant, the system (eq. (\ref{syst lin})) reduces to the single equation
\begin{equation}
\sigma_q= \frac{C_{q,0}}{\left(\left(\frac{\lambda}{\lambda_q}\right)^2 -1-\Sigma_{q,q} \right)},  
\label{syst lin 1 mode}
\end{equation}  
from which the eigen wavelength $\lambda_q^{pert}$ of the mode of the studied structure  can be deduced (this is the pole of $\sigma_q$)
\begin{equation}
\lambda_q^{pert} = \lambda_q  \sqrt{ 1+ \Sigma_{q,q} },
\label{pole q}
\end{equation}  
and appears as a modification of the eigen wavelength of the excited eigen mode caused by the second order self-coupling of the mode through the diffraction orders.  
Let us note that the imaginary part of this pole gives the  half-width of the resonance peak. As $\lambda_q$ is real, the width of the peak depends on the second order coupling coefficients $C_{q,n,q}$, and depicts the  leakage in the substrate and the superstrate of the mode excited. As we consider that the zero diffraction order is the only propagative one, we can expect that the width of the resonance peak is mainly given by $C_{q,0,q}$, and that $C_{q,n,q}$ for $n\neq0$ plays a minor role. $C_{q,0,q}$ includes the $q^{\text{th}}$ harmonic of the permittivity of the grating, as it was already underlined in the literature concerning guided mode resonance gratings \cite{Norton_JOSAA_1997,Evenor_EurPhysJD_2012,Paddon_OptLett_1998,Lemarchand_OptLett_1998}. 

Further, an important property can be easily derived from the expression of $\mathbf{R}^{pert}_q$ (see eq. (\ref{Rpert})), which is valid even when several orders are resonant.   The determinant of $\mathbf{R}^{pert}_q$ is null, hence one eigenvalue of $\mathbf{R}^{pert}_q$ is null, and the other, called $\mu_q$, is equal to the trace of $\mathbf{R}^{pert}_q$:
\begin{equation}
\mu_q=\sigma^s_q C^s_{0,q}(0)+ \sigma^p_q C^p_{0,q}(0).
\label{muq}
\end{equation}
 The eigenvector associated with $\mu_q$ is $\bvec{V}= [C^s_{0,q}(0) ; C^p_{0,q}(0)]$. The eigenvector associated with the null eigenvalue is $\bvec{V}_0= [\sigma_q^p ; -\sigma_q^s]$. Similar properties can be derived for the transmission matrix. 

Now, in the particular case where only one order $q$ is resonant, the non-null eigenvalue of $\mathbf{R}^{pert}_q$ takes the form 
\begin{equation}
\mu_q=\frac{\lambda_q^2(C^s_{q,0} C^s_{0,q}(0)+ C^p_{q,0} C^p_{0,q}(0))}{\lambda^2-(\lambda_q^{pert})^2},
\label{muq one}
\end{equation}
where $C^s_{q,0}$ and  $C^p_{q,0}$ are obtained  by considering respectively a $s$ and a $p$ incident field in $C_{q,0}$ (eq. (\ref{def Cq0})).
 In this case, the eigenvector $\bvec{V}_0$ associated to the null eigenvalue is collinear to $[C^p_{0,q}(0); - C^s_{0,q}(0)]$, and is hence orthogonal to  $\bvec{V}$. This means that the mode is fully excited with the incident polarization corresponding to $\bvec{V}$, and not at all with the orthogonal polarization, as already observed in \cite{Fehrembach_JOSAA_2002}.  The numerator $C^s_{q,0} C^s_{0,q}(0)+ C^p_{q,0} C^p_{0,q}(0)$ appearing in eq. (\ref{muq one}) depicts the coupling in and out of the mode when the structure is illuminated with the suitable incident polarization. 
Note that this polarization may not be the one giving the maximum reflectivity, since the reflectivity of the reference structure must also be taken into account (see eq. (\ref{Rref+Rpert}), and the right bottom sketch on Fig. \ref{fig resume}). 

\subsection{Situation with two resonant orders}

In the situation where two diffraction orders $q$ and $q'$ are resonant, the system eqs. (\ref{syst lin}) reduces to two coupled equations  
\begin{equation}
\begin{bmatrix}
\frac{\lambda^2-(\lambda_q^{pert})^2}{(\lambda_q)^2} & -( C_{q,q'} +\Sigma_{q,q'}) \\
 -( C_{q',q} + \Sigma_{q',q})  & \frac{\lambda^2-(\lambda_{q'}^{pert})^2}{(\lambda_{q'})^2} 
 \end{bmatrix}
\begin{pmatrix}
\sigma_q \\ \sigma_{q'}
 \end{pmatrix}
 = 
 \begin{pmatrix}
C_{q,0} \\ C_{q',0}
 \end{pmatrix},
 \label{syst lin 2}
 \end{equation}
where $\lambda_q^{pert}$ and $\lambda_{q'}^{pert}$ are given by eq. (\ref{pole q}) and correspond to the eigen wavelengths of the modes $q$ and $q'$ when their mutual coupling is not taken into account. The eigen wavelengths of the  studied structure   are the wavelengths for which the determinant of the system of eqs. (\ref{syst lin 2}) is null. They are split on each side of the wavelength given by $\sqrt{[(\lambda_q^{pert})^{2}+(\lambda_{q'}^{pert})^{2}]/2}$, the splitting being governed by the coupling between the modes (anti-diagonal terms in eq. (\ref{syst lin 2})), as already shown in \cite{Norton_JOSAA_1997,Evenor_EurPhysJD_2012,Paddon_OptLett_1998,Lemarchand_OptLett_1998}.

 \begin{figure}[b]
\includegraphics[width=4cm]{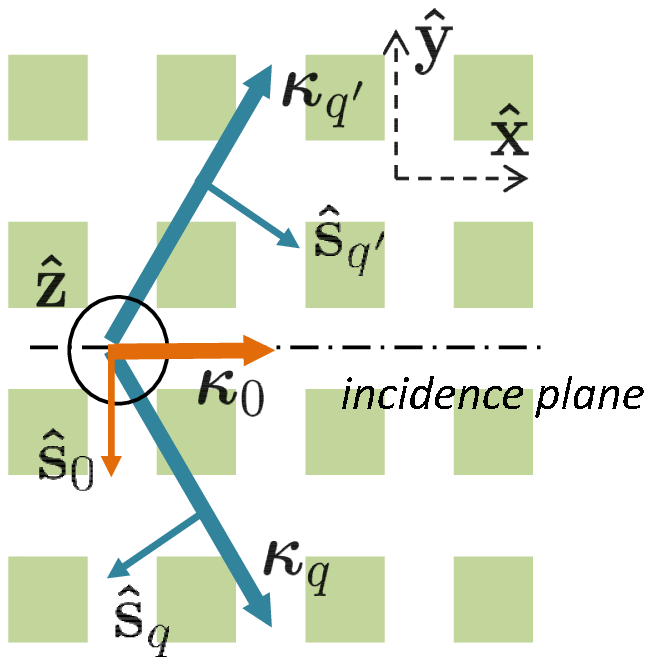}
\caption{Configuration where the plane of incidence is a plane of symmetry of the structure.}
\label{fig incidence sym}
\end{figure}

In the particular case where the plane of incidence is a plane of symmetry of the structure (see Fig. \ref{fig incidence sym}), the two diffraction orders $q$ and $q'$ excite one guided mode of the reference structure, along two directions. Thus, the following relations  are valid (both for TE modes and TM modes): $\lambda_q= \lambda_{q'}$; $C_{q,q'}=C_{q',q}$; $\Sigma_{q,q'}=\Sigma_{q',q}$; $\Sigma_{q,q}=\Sigma_{q',q'}$. 
Moreover, in the case of a TE mode, we have $C^s_{q,0}=C^s_{q',0}$,  $C^p_{q,0}=-C^p_{q',0}$, because $\buvec{s}_q$ and $\buvec{s}_{q'}$ are symmetrical with respect to the plane ($\buvec{s}_0,\buvec{z}$), and anti-symmetrical with respect to the plane  ($\buvec{\veckappa}_0,\buvec{z}$)  (see Fig. \ref{fig incidence sym}). In the same manner, for a TM mode, $C^s_{q,0}=-C^s_{q',0}$,  $C^p_{q,0}=C^p_{q',0}$.

The eigen wavelengths $\lambda_{qq',\pm}^{pert}$ of the two modes of the studied structure are given by $(\lambda_{qq',\pm}^{pert})^2=(\lambda_q)^2 \left[1 + \Sigma_{q,q}  \pm (C_{q,q'}  + \Sigma_{q,q'}\right)]$. Note that in the particular case of normal incidence, we have  $\Sigma_{q,q}=-\Sigma_{q,q'}$ for a TE mode and $\Sigma_{q,q}=\Sigma_{q,q'}$ for a TM mode, so that one of the eigen wavelengths is real, which corresponds to the anti-symmetric mode which can not be excited by a symmetric normal incident plane wave. 
Then, the  coefficients $\sigma_q$ and $\sigma_q'$ are deduced from eq. (\ref{syst lin 2}): 
$\sigma_q=\frac{(\lambda^2-(\lambda_q^{pert})^2)C_{q,0} +( C_{q,q'} + \Sigma_{q,q'})C_{q',0}} {(\lambda_q)^2(\lambda^2-(\lambda_{qq',+}^{pert})^2)(\lambda^2-(\lambda_{qq',-}^{pert})^2)}$, with a similar expression for $\sigma_q'$ obtained by exchanging $q$ and $q'$. From the relation between $C_{q,0}$ and $C_{q',0}$ we deduce that for a TE mode $\sigma_q^s=\sigma_{q'}^s$ and $\sigma_q^p=-\sigma_{q'}^p$ and for a TM mode $\sigma_q^s=-\sigma_{q'}^s$ and $\sigma_q^p=\sigma_{q'}^p$. Now, using the expression of $\mathbf{R}^{pert}_q$ (eq. (\ref{Rpert})) we obtain 
\begin{equation}
\mathbf{R}^{pert}_q+\mathbf{R}^{pert}_{q'}=
\begin{bmatrix}
2 \sigma_q^sC^s_{0,q} & 0 \\ 0 & 2 \sigma_q^pC^p_{0,q}
\end{bmatrix},
\end{equation}
both for a TE mode and a TM mode. This means that the coupling between the two excited guided modes generates two hybrid modes, one of which is excited with a $s$ polarization, while the other is excited with a $p$ polarization, thus confirming the observations reported in the literature \cite{Mizutani_JOSAA_2001,Fehrembach_JOSAA_2003,Lacour_JOSAA_2003,Niederer_OptExpr_2005}. A full vectorial analysis was necessary to depict this phenomenon.

%\begin{equation}
%\lambda_{qq',\pm}^{pert}=\frac{(\lambda_q^{pert})^2+(\lambda_{q'}^{pert})^2}{2} \pm %\sqrt{(\left(\frac{(\lambda_q^{pert})^2-(\lambda_{q'}^{pert})^2}{2} \right)^2+(\lambda_{q}\lambda_{q'})^2( C_{q,q'} + \sum_{n \notin %\mathcal{Q}}   C_{q,n,q'} )( C_{q',q} + \sum_{n \notin \mathcal{Q}}   C_{q',n,q} )}.
%\end{equation}
%We find the well known result that the 
% and  $\sum_{n \notin \mathcal{Q}} C_{q,n,q}=-\sum_{n \notin \mathcal{Q}} C_{q',n,q'}$

%%%%%%%%%%%%%%%%% SECTION 3
\section{Numerical verification}
For the validation of the method, we compare the results to the reflection and transmission coefficients calculated with an home made code based on the Fourier Modal Method \cite{Li_JOSAA_1997}. 
Several configurations are considered, involving one or several coupled modes, TE or TM mode, and 1D or 2D gratings. We tested the convergence with respect to the number $N$ of coefficients taken in the sum  
on the non-resonant orders of eq. (\ref{syst lin}) ($\Sigma_{q,q'}=\sum_{n (\notin \mathcal{Q})=-N}^{N}   C_{q,n,q'} $). We also tested the validity of the method with respect to the depth $h$ of the grating.

%\section{\label{sec:level1}First-level heading}
% sections are not used for PRL papers

\subsection{Configuration 1: TM mode, 1D grating, conical incidence}

The structure of configuration 1 is composed with a substrate with a dielectric permittivity of 2.097, a first layer with a 301.2\,nm thickness and a 4.285 permittivity, a second layer with a 140.4\,nm thickness and a 2.161 permittivity, a grating with depth 70\,nm, period 838\,nm, grooves width 300\,nm engraved in a 2.161 permittivity material and filled with air (permittivity 1). The superstrate is also air. The angles of incidence are $\theta=15\degree$ and $\phi=50,5$\degree  (see Fig.\ref{fig configuration} for the definition of $\theta$ and $\phi$). In this configuration, a TM guided mode is excited around 1.459\,$\mu$m, and under conical incidence through the -1 diffraction order. The resonance is observable for both $s$  (incident electric field perpendicular to the plane of incidence) and $p$ polarizations  (incident magnetic field perpendicular to the plane of incidence) . 

\begin{figure}
\begin{tabular}{c}
\includegraphics[width=8cm]{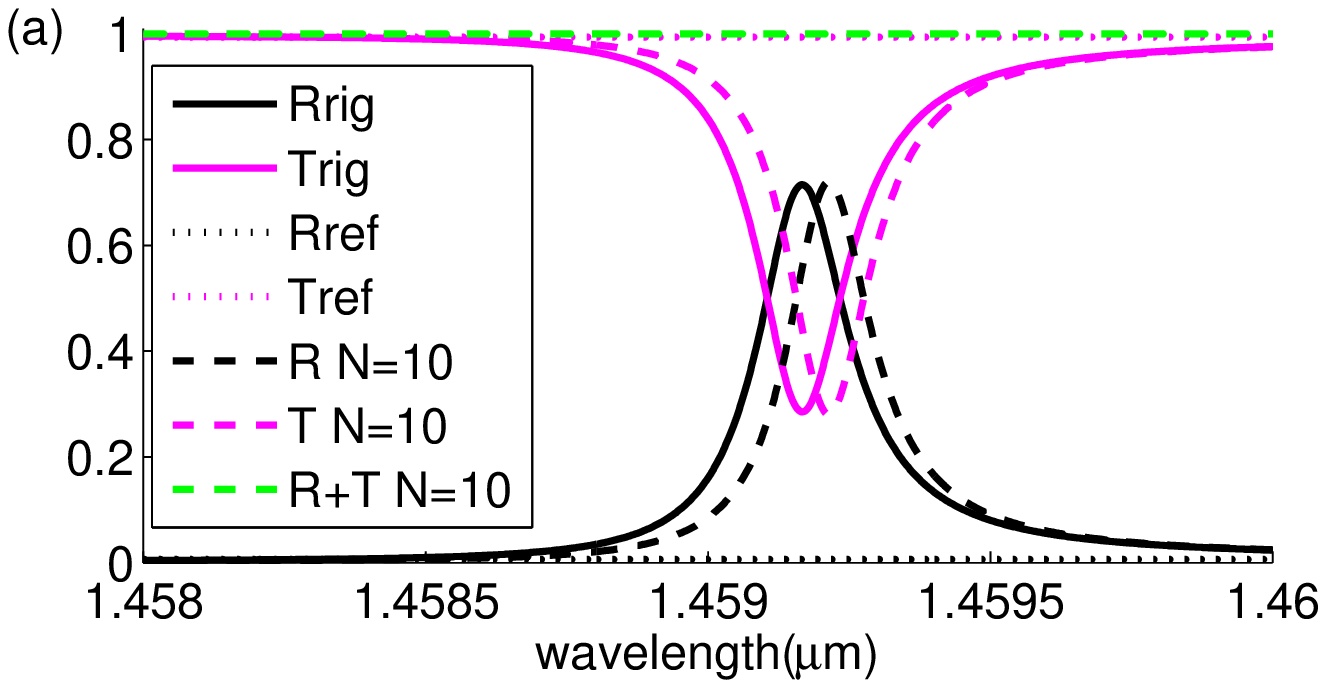} \\ \includegraphics[width=8cm]{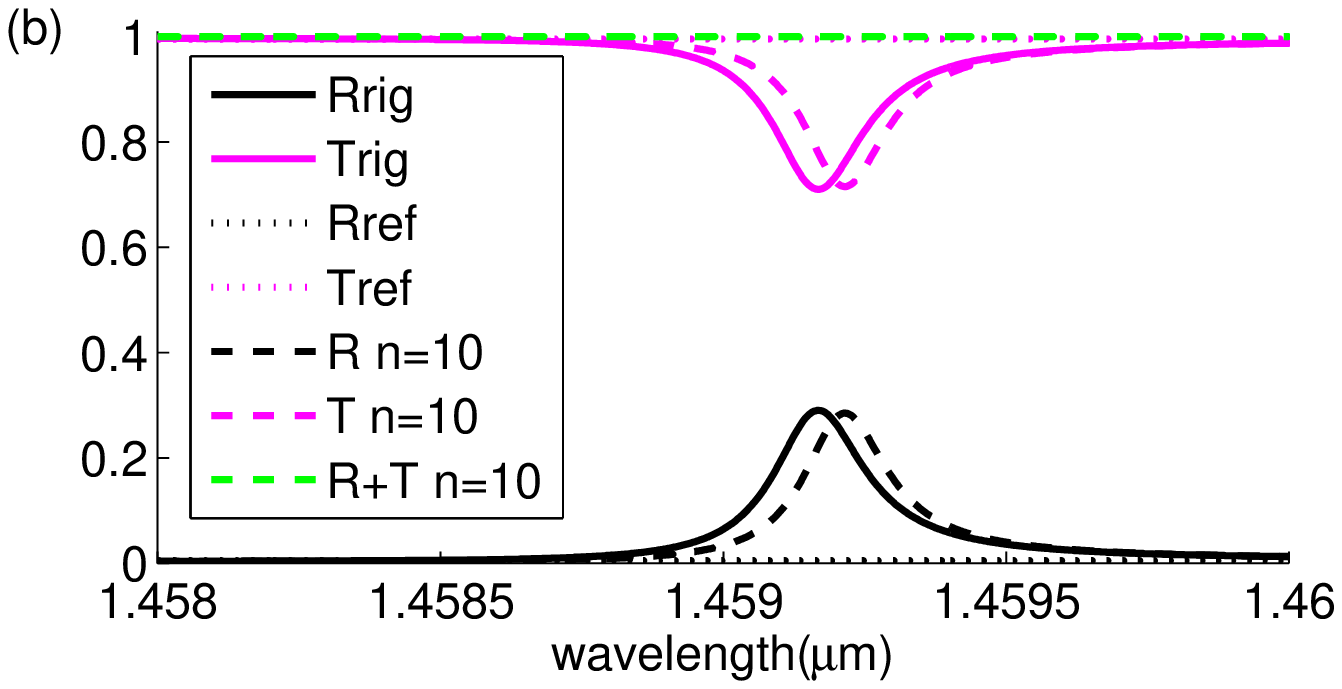} \\ \includegraphics[width=8cm]{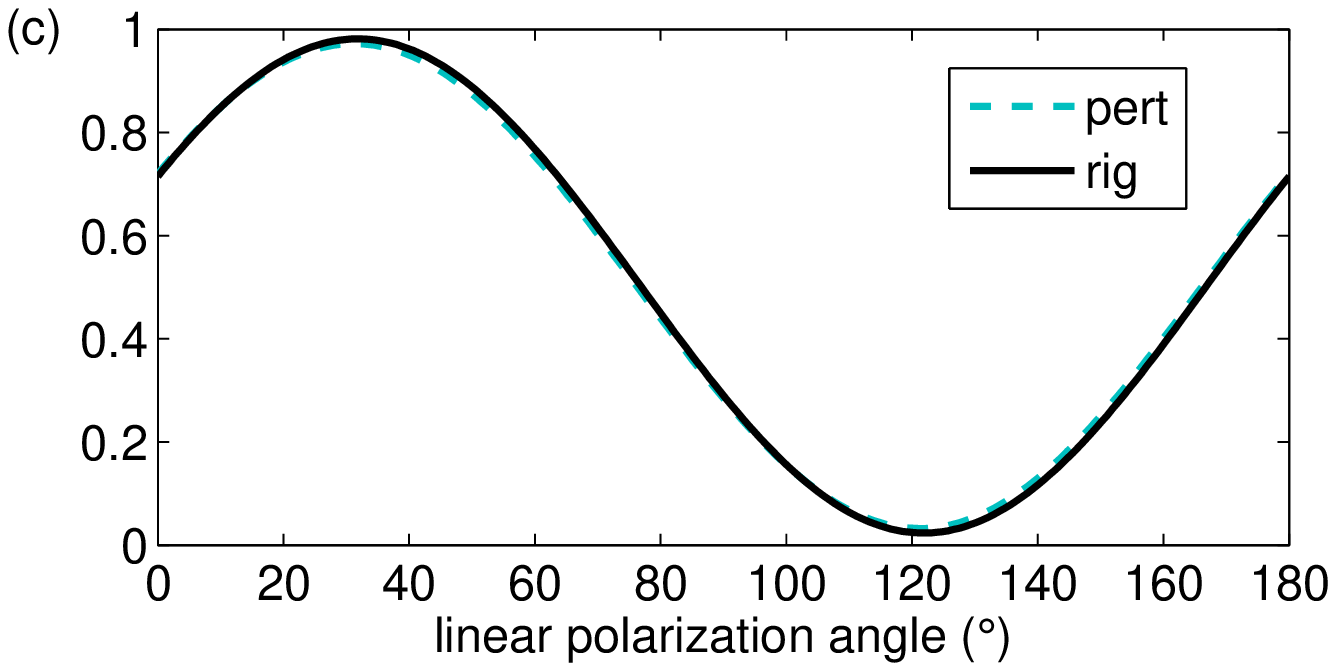}
\end{tabular}
\caption{\label{fig:fig1} Configuration 1 - Reflectivity and transmittivity spectra calculated rigorously (straight lines, Rrig and Trig), with the approached method  for $N=10$ (dashed lines, R and T, and sum R+T), and for the reference structure (dotted lines, Rref and Tref):  (a) for the $s$ incident polarization and (b) for the $p$ polarization. (c) Reflectivity at resonance for any linear polarization with respect to the angle between the electric field and the $s$ polarization, calculated with the approached method for $N=10$ (dotted cyan line) and rigorously (black straight line).}
\end{figure}

We compare in Fig.\ref{fig:fig1} ((a) for $s$ incident polarization and (b) for $p$ polarization) the reflectivity and transmittivity calculated with the approached method for $N=10$ (dashed lines, R and T) and with the rigorous method (solid lines, Rrig and Trig). We also plot the reflectivity and transmittivity for the reference planar structure (dotted lines, Rref and Tref) and the sum of the reflectivity and transmittivity calculated with the approached method (dashed green line, R+T). 

First, we observe that the resonance is well depicted, with a resonance wavelength, a width and maxima close to the rigorous ones, both for $s$ and $p$ polarizations, for the transmission and the reflection. We also observe that R+T is close to 1, i.e. the energy conservation is satisfied. Last, we observe that the reflectivity and transmittivity of the studied structure tend to that of the unperturbed structure far from the resonance. 

In Fig. \ref{fig:fig1}(c), we plot the reflectivity at resonance for any linear polarization with respect to the angle between the electric incident field and the $s$ polarization, both for the rigorous and the approached calculation. As expected from the analysis of the previous paragraph in the case of one resonant order only, we have a polarization (quasi-linear polarization with an angle 31.7$\degree$ with the $s$ polarization) for which the mode is fully excited, and not at all for the orthogonal polarization (quasi-linear polarization with an angle 121.7$\degree$). 

\begin{figure}
\begin{tabular}{c}
\includegraphics[width=8cm]{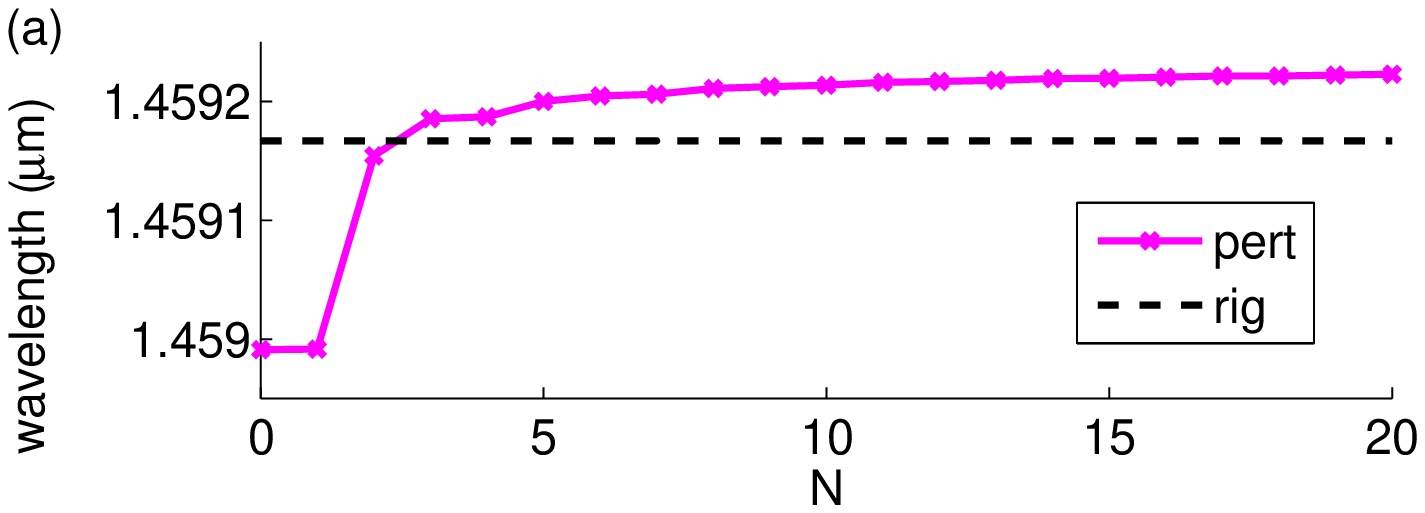} \\ \includegraphics[width=8cm]{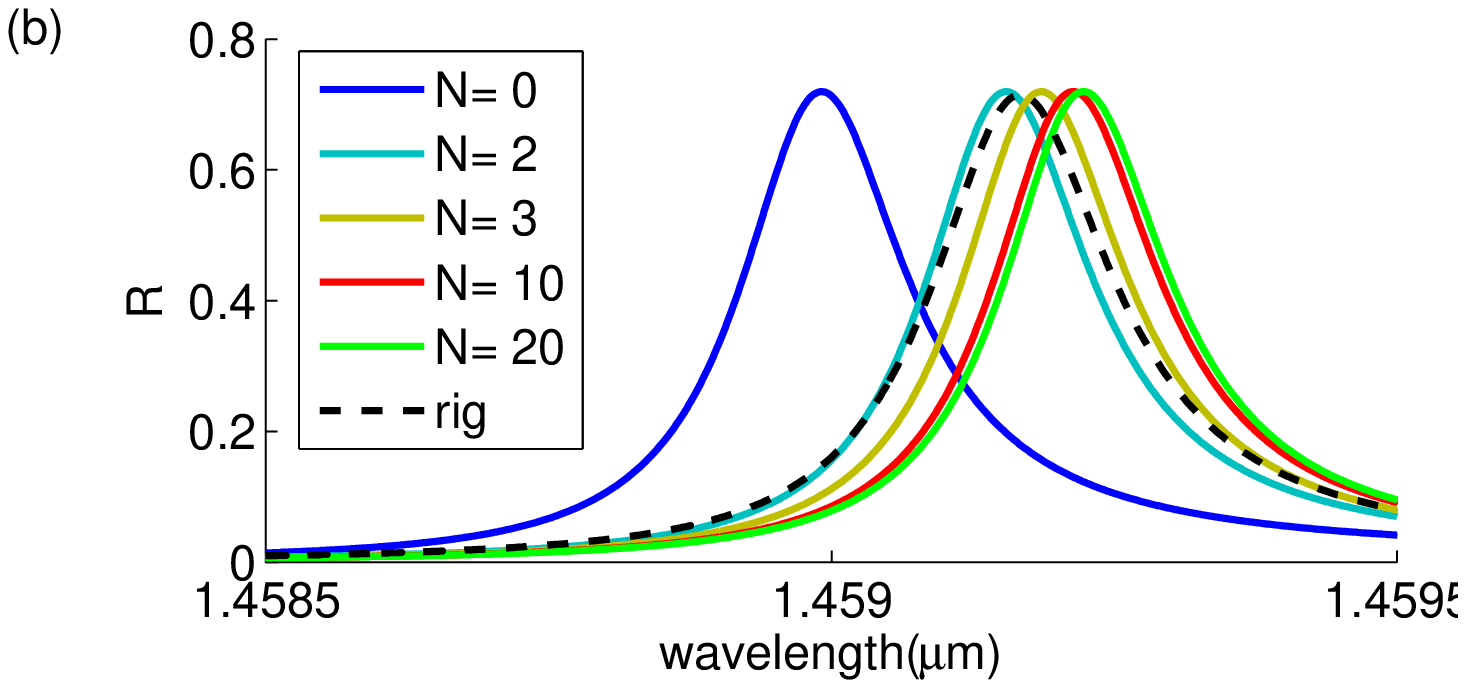}
\end{tabular}
\caption{\label{fig:fig2} Configuration 1 - Convergence of the position of the resonance peak with respect to N. (a) Resonance wavelength calculated with the approached model with respect to N (solid line with stars) and  calculated rigorously (dashed line). (b) Spectrum calculated with the approached model for various values of N (solid lines), and rigorously (dashed line). }
\end{figure}

To study the impact of the number of orders taken when summing the coefficients $C_{q,n,q'}$ for $n$ from $-N$ to $N$, we plot in Fig. \ref{fig:fig2}(a) the resonance wavelength calculated with the approached model for various values of $N$, and the resonance wavelength calculated rigorously, for comparison. The spectra for some values of N are plotted in Fig.\ref{fig:fig2}(b). We observe that the peak calculated with the approached method is positioned at a shorter wavelength than  the peak calculated rigorously when $N=0$, and it moves to higher wavelengths  when $N$ grows. The final difference is no more than a quarter of the bandwidth of the peak. As expected, the width of the peak is not much modified with respect to that obtained for $N=0$.

We also considered a case where a TE mode is excited under conical incidence. We found that the resonance wavelength difference between the approached and rigorous calculations  converges to one bandwidth over 6 (not shown). 
%This can be explained by noting that the excited guided modes  propagate in directions close to the directions of periodicity of the grating (due to the coupling condition). And our reference structure, with the geometric average of the permittivity in the $(x,y)$ plane, is more suitable for TE modes  than for TM modes for which the harmonic average is required. 
%One could expect that the reference structure chosen (grating replaced with an homogneous anisotropic material with permittivity equals to the geometric average of the grating permittivity in the ($x,y$) plane and to the harmonic average along $z$) is more suitable to model 2D gratings (with $\pi/2$ rotation invariance around $z$) that 1D gratings, as the 1D grating creates a stronger form anisotropy in the $(x,y)$ plane, than the 2D grating. Yet, it is important to note that the guided modes excited propagate in directions close to the directions of periodicity of the grating (due to the coupling condition). As a consequence, the fact that the 1D grating has a translation invariance along its ridges has a  minor impact on the propagation of the guided mode.

\subsection{Configuration 2: TM mode, 1D grating, quasi-normal classical incidence}

We now consider the quasi-normal incidence case where a guided mode can be excited along two counter propagative directions through two opposite diffraction orders. The grating is 1D and the plane of incidence is perpendicular to the grating grooves. The combination of the two modes gives one mode with a field symmetric and another with a field anti-symmetric with respect to the plane normal to the direction of propagation of the modes. They correspond to the edges of a band gap in the dispersion relation of the structure. The symmetric mode is well excited with an incident plane wave, and gives a broad peak, while the anti-symmetric mode is scarcely excited leading to a thin peak that disappears under normal incidence.
The considered structure is the same as in the previous paragraph, the only difference being the incident field. The angles of incidence are set to $\theta=0.01\degree$ and $\phi=0$\degree. In this configuration, a TM guided mode is excited around 1.357\,$\mu$m through the (+1) and (-1) diffraction orders of the grating, and  the resonances are observable for the $p$ polarization. 

\begin{figure}
\begin{tabular}{c}
\includegraphics[width=8cm]{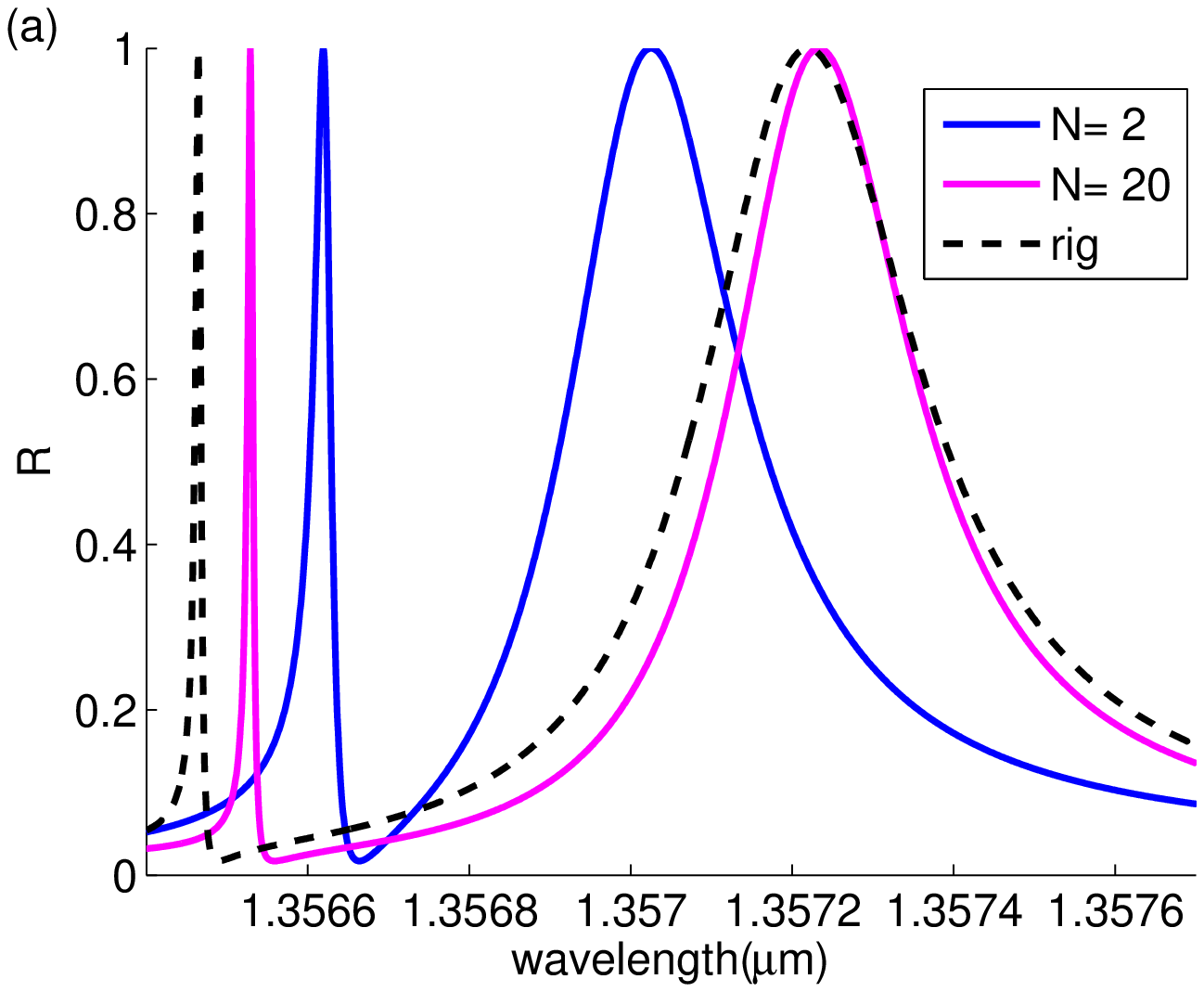} \\ \includegraphics[width=8cm]{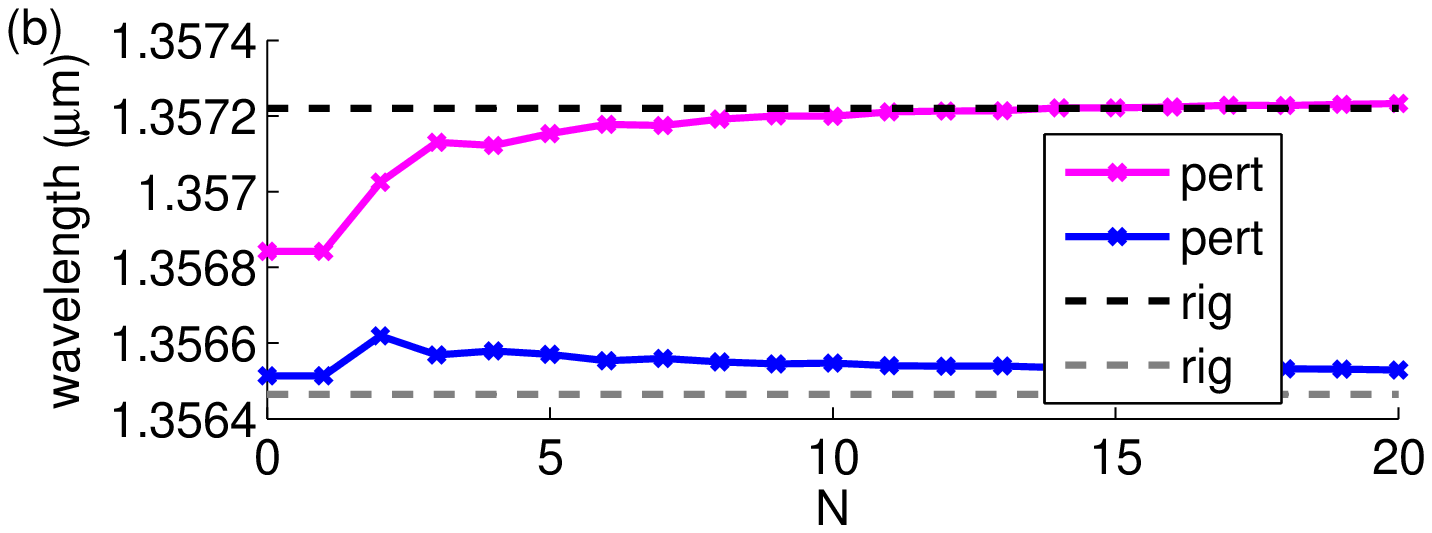}  \\ \includegraphics[width=8cm]{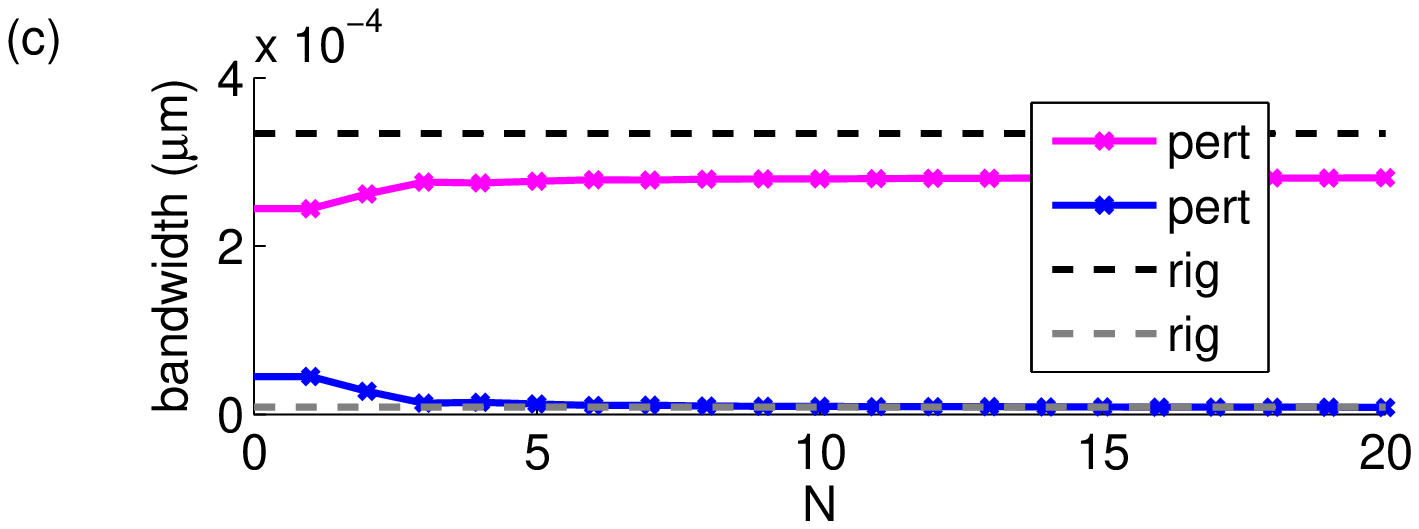}
\end{tabular}
\caption{\label{fig:fig3} Configuration 2 - Convergence of the position of the resonance peaks with respect to N. (a) Spectrum calculated for $p$ incident polarization for $N=2$ (blue solid line), $N=20$ (pink solid line) and rigorously (dashed line). (b) Resonance wavelength and (c) bandwidth of the two peaks calculated with the approached model with respect to N (solid lines with stars) and  calculated rigorously (dashed lines).}
\end{figure}

We plot in Fig. \ref{fig:fig3}(a) the spectrum calculated with the approached model for $N=2$, $N=20$ and rigorously. As expected, we observe a broad and a thin peak. The approached calculation gives the right  layout of the two peaks, with the broad peak for upper wavelengths and the thin peak for smaller wavelengths. Taking into account more $C_{q,n,q'}$ coefficients brings the approached calculation closer to the rigorous one. We have checked that the energy conservation (R+T=1) is fulfilled also in this case (not shown on the curves).

The change of the resonance wavelength and bandwidth with respect to $N$ can be seen in Fig. \ref{fig:fig3}(b) and (c). The position of the two peaks is related to the coupling between the two counter propagative modes: the higher the coupling, the greater the difference between the two resonance wavelengths. Hence, we observe in Fig. \ref{fig:fig3}(b) that the distance between the two peaks increases with the number of coupling integrals  $C_{q,n,q'}$. The separation of the two peaks has an impact on the shape, and as a consequence, on the width of the peaks (see Fig.  \ref{fig:fig3}(c)).

We also considered a case where a TE mode is excited under quasi-normal incidence (not shown here). The spectrum obtained with the approached model are remarkably close to the rigorous results (closer than for the TM mode).

\subsection{Configuration 3: TE mode, 2D grating, oblique incidence}

\begin{figure}
\begin{tabular}{c}
\includegraphics[width=8cm]{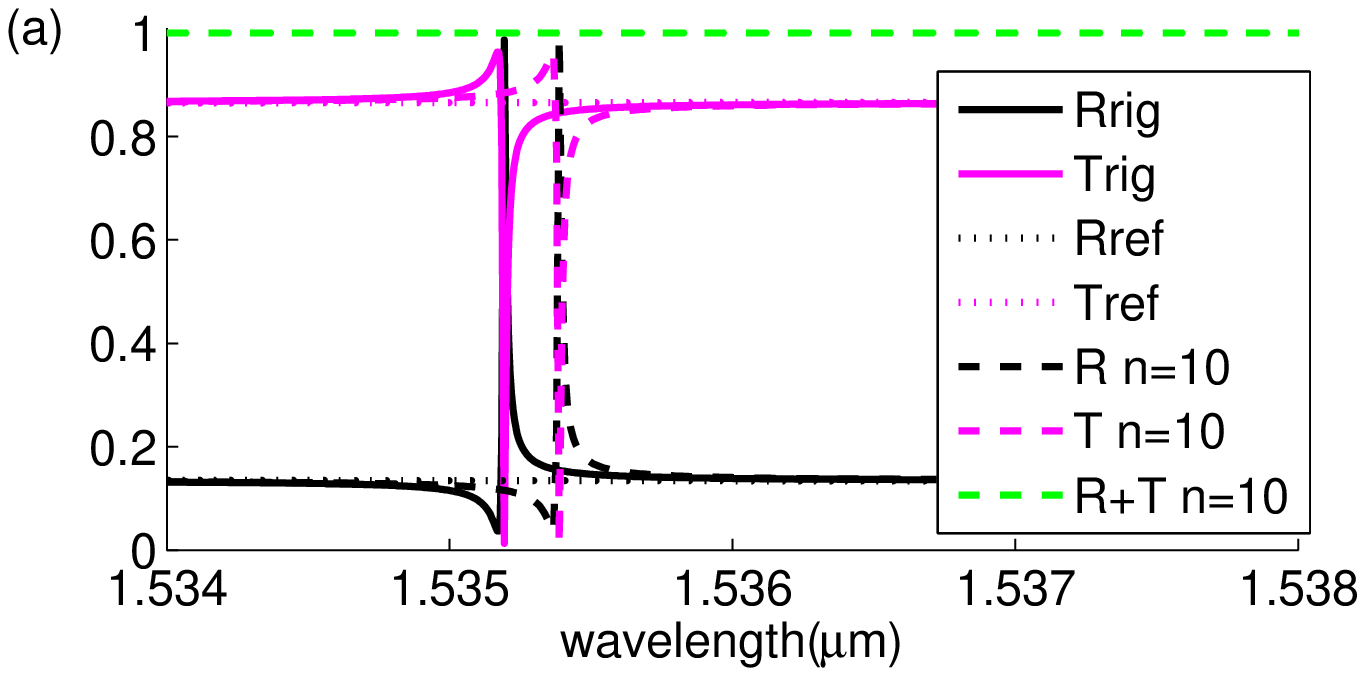} \\ \includegraphics[width=8cm]{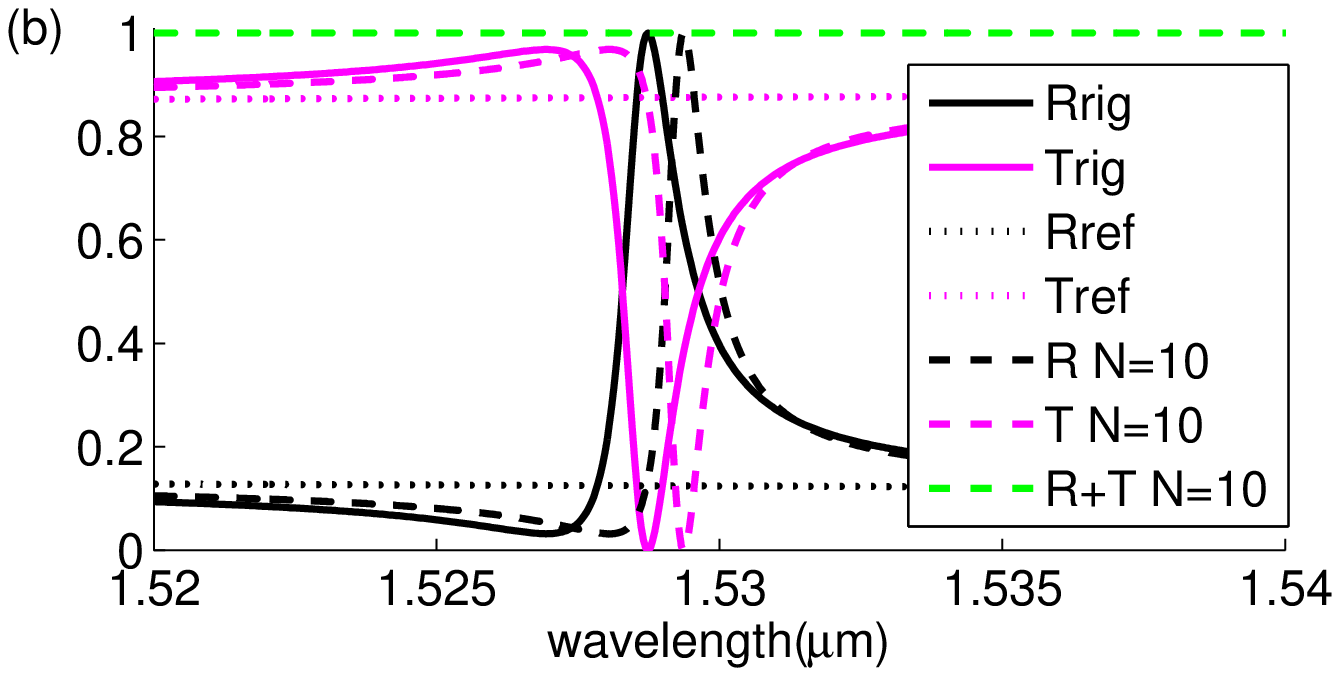}
\end{tabular}
\caption{\label{fig:fig4} Configuration 3 - Reflectivity and transmittivity spectra calculated rigorously (straight lines, Rrig and Trig), with the approached method (dashed lines, R and T, and sum R+T) for $N=10$, and for the reference structure (dotted lines, Rref and Tref).  (a) For the $s$ incident polarization, (b) for the $p$ polarization.}
\end{figure}

Our third example is a 2D square grating illuminated under oblique incidence along one direction of periodicity (see Fig.\ref{fig incidence sym}). The structure is composed with a substrate with dielectric relative permittivity 2.25, a layer with thickness 400\,nm and relative permittivity 4.0, and a 2D square grating with period 870\,nm both along $x$ and $y$, made with square holes with  300\,nm width, 250\,nm depth engraved in a material with relative permittivity 4.0. The angles of incidence are $\theta=13\degree$ and $\phi=0\degree$. In this configuration, a TE mode can be excited through the (0,-1) and (0,+1) diffraction orders around 1.53\,$\mu$m. The simultaneous excitation of a guided mode in the two symmetrical directions generates two modes, one with a field symmetric, and the other anti-symmetric with respect to the $(x,z)$ plane. As shown in the previous section (case of two resonant orders), the symmetric mode can be excited with a $p$ incident polarization, while the anti-symmetric mode can be excited with a $s$ polarization.  The spectrum calculated with the approached model for $N=10$ (dashed lines, R and T), with the rigorous numerical method (solid lines,   Rrig and Trig), and for the reference structure (dotted lines, Rref and Tref) are plotted in Fig. \ref{fig:fig4} for a $s$ incident polarization (a) and a $p$ polarization (b). Again, the two peaks are well represented by the approached model, for the bandwidth, maximum and minimum. They are shifted a little toward greater wavelengths with respect to the rigorous peak. The energy conservation is fulfilled.

%One could expect that the reference structure chosen (grating replaced with an homogneous anisotropic material with permittivity equals to the geometric average of the grating permittivity in the ($x,y$) plane and to the harmonic average along $z$) is more suitable to model 2D gratings (with $\pi/2$ rotation invariance around $z$) that 1D gratings, as the 1D grating creates a stronger form anisotropy in the $(x,y)$ plane, than the 2D grating. Yet, it is important to note that the guided modes excited propagate in directions close to the directions of periodicity of the grating (due to the coupling condition). As a consequence, the fact that the 1D grating has a translation invariance along its ridges has a  minor impact on the propagation of the guided mode.

\subsection{Configuration 4 : TE mode, 1D grating, classical incidence - variation of the grating depth $h$}

Our fourth example is a 1D grating illuminated under oblique classical incidence. The angles of incidence are set to $\theta=30\degree$ and $\phi=0\degree$.    The structure is composed with a substrate with dielectric relative permittivity 2.25, a layer with thickness 250\,nm and relative permittivity 4.0, and a 1D grating with period 742.2\,nm and 442.2\,nm groove width, engraved in a material with relative permittivity 4.0. We are interested in the resonance due to the excitation of a TE mode around 1.57$\,\mu$m and to its evolution when the depth of the grating is varying.

\begin{figure}
\begin{tabular}{c}
\includegraphics[width=8cm]{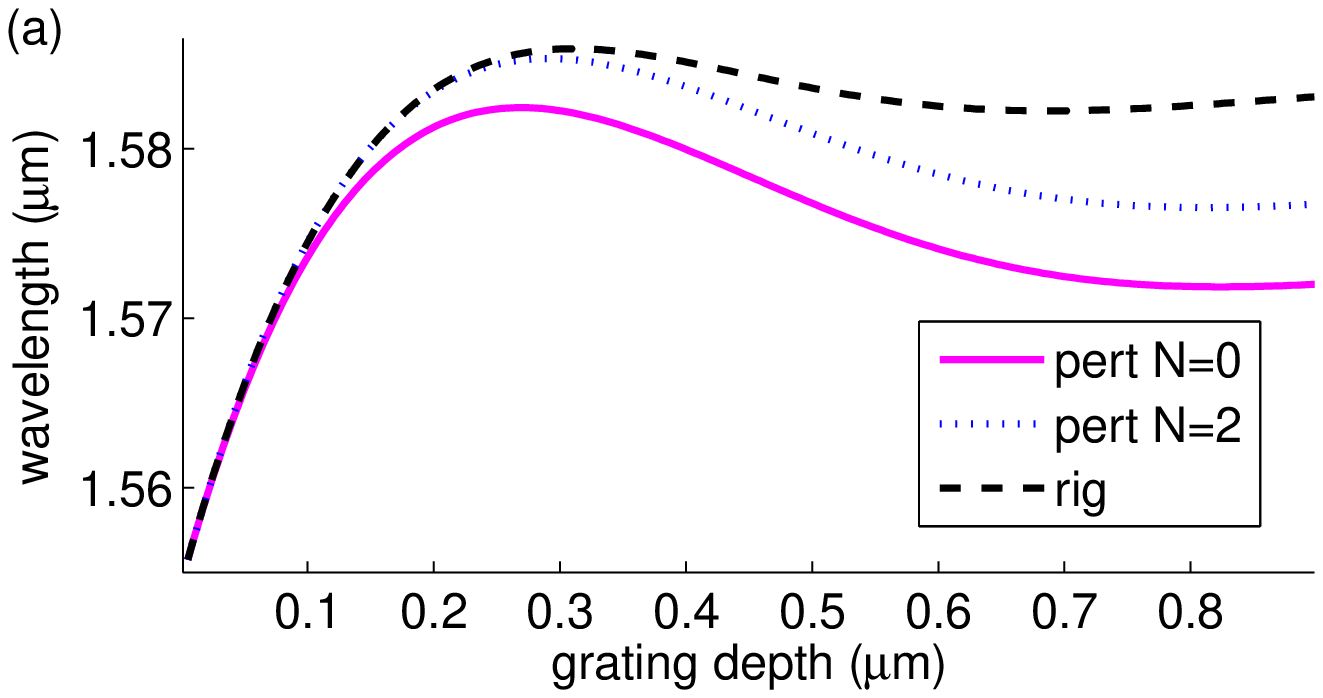} \\ \includegraphics[width=8cm]{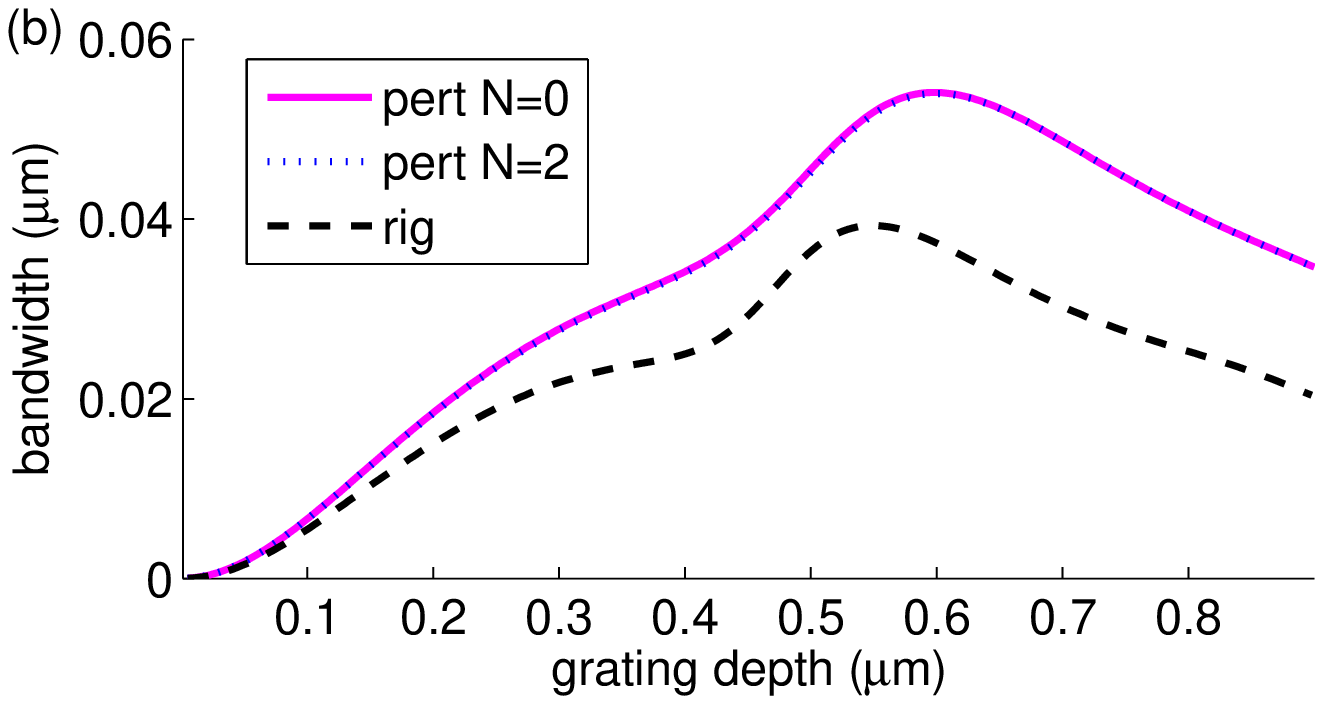}
\end{tabular}
\caption{\label{fig:fig5} (a) Configuration 4 - Resonance wavelength calculated rigorously (dashed line), and with the approached method for $N=0$ and $N=2$ (solid pink and dashed blue lines). (b) Bandwidth calculated rigorously (dashed line), and with the approached method for $N=0$ and $N=2$ (solid pink and dashed blue lines). }
\end{figure}

We plot in Fig.\ref{fig:fig5} the centering wavelength (a) and the width (b) of the peak obtained with the rigorous numerical code and the approached method (for $N=0$ and $N=2$), with respect to the depth of the grating (from 0 to 1$\,\mu$m). First, we observe that the global shape of the curves is similar for the rigorous and the approached methods: the resonance wavelength reaches a maximum for a grating depth around 300\,nm, and the bandwidth for a grating depth between 500\,nm and 600\,nm. Second, the resonance wavelength is very well calculated with the approached method up to $\lambda/15$ for $N=0$ and $\lambda/6$ for $N=2$. It is all the more impressive that the relative permittivity of the material in which the grating is engraved is 4.0, which is not small. This relatively good robustness of the method concerning the resonance wavelength with respect to the grating depth may be attributed to the evanescent behavior of the mode in the grating layer. Last, we observe that the bandwidth is over estimated with the approached method, and that increasing $N$ does not improve the description of the bandwidth, as expected.

%%%%%%%%%%%%%%%%%%%% CONCLUSION
\section{Conclusion}

A full vectorial approached model has been proposed to describe the reflectivity and transmittivity properties of guided mode resonance gratings. We showed how the reflectivity (respectively transmittivity) matrix can be expressed as the sum of a resonant and a non-resonant term.
The non-resonant term is the reflectivity (respectively transmittivity) matrix of a  planar reference structure. The resonant term is a sum of matrices (one for each mode excited), each matrix being  expressed with coupling integrals involving the modes of the planar reference structure and the radiative modes. Our model is of course valid for the scalar configuration (typically a 1D grating illuminated along its direction of periodicity), and able to depict the physical behaviors  already mentioned in the literature in this configuration.

Moreover, we demonstrate additional properties in the vectorial configuration (typically a 1D grating illuminated under conical incidence, or a 2D grating). A fundamental property of our resonant matrix is that one of its eigenvalue is null, the other eigenvalue being  resonant. The eigenvector associated with the non-null eigenvalue corresponds to the polarization for which the eigen mode is fully excited. Furthermore, writing the reflectivity (or transmittivity) matrix as the sum of a non-resonant and  a resonant matrix allows the identification, in the polarization of the reflected (or transmitted) field, of the influence of the non-resonant field and that of  the resonant field. 
We believe that this model  provides a physical insight especially for configurations where the polarization of the modes is not trivial (not $s$ or $p$ polarization), and where the addition of the non-resonant  and the resonant terms leads to a polarization that differs from the polarization expected from the excited mode alone.   

We validated our model in various configurations (with TE or TM modes, one resonant or two resonant orders, 1D or 2D gratings), and shown a good robustness with respect to the grating depth.

Our model can be easily extended to configurations where several gratings are included inside the stack, and where the materials are anisotropic with $z$ axis symmetry. For bi-anisotropic materials, there is no technique to express the Green's tensor as two independent scalar Green's functions. One interesting further development of our method would be to consider configurations that are not periodic, as for example a coupling grating, or a Cavity Resonator Integrated Grating Filter (CRIGF, guided mode resonance grating surrounded by Bragg reflectors) \cite{Ura_ICTON_2011}. In this case, as the spacial frequencies in the Fourier space are no more discrete but continuous, the problem can not be expressed as  a set of coupled equations, which requires further investigations.

\section{Appendix}

 \subsection{The vectorial problem for a planar structure expressed as two scalar problems}

The planar reference structure has a relative permittivity which is anisotropic with symmetry axis $z$. Thus, the diffraction or homogeneous vectorial problems associated with the reference problem  can be divided into two scalar problems corresponding to the transverse electric and transverse magnetic cases (transverse with respect to the direction of propagation of the mode for the homogeneous problem and to the plane of incidence for the diffraction problem).  

The equation satisfied by the electric field $\Evec^{ref}_m(z)$ for the planar reference structure is (see eq. (\ref{Helmholtz reference}))
\begin{equation}
  \opOmega_m(\Evec^{ref}_m(z))-k_0^2  \tenseps^{ref}(z) \Evec^{ref}_m(z)=\bvec{0}, 
  \label{Helmholtz reference appendix}
  \end{equation}
  where the expression of the operator $\opOmega_m$ is given by eq. (\ref{opOmega}). 
  
  To each diffraction order $m$ with in-plane wavevector $\bvec{\veckappa}_m$, it is possible to associate an orthonormal basis $(\buvec{s}_m,\buvec{\veckappa}_m,\buvec{z})$ (even for evanescent orders), where $\buvec{s}_m=\buvec{\veckappa}_m \times \buvec{z}$.
%\begin{equation}
%\buvec{u}_m^3=\buvec{z}, \quad \quad \buvec{\veckappa}_m=\veckappa_m/\kappa_m, \quad  \quad \buvec{s}_m=\buvec{\veckappa}_m %\times \buvec{u}_m^3.
%\end{equation}
The operator  $\opOmega_m$ can then be written as
  \begin{equation}
  \opOmega_m=(\kappa_m^2-\partial_z^2)\buvec{s}_m\buvec{s}_m-\partial_z^2\buvec{\veckappa}_m\buvec{\veckappa}_m+\kappa_m^2\buvec{z}\buvec{z}+i\kappa_m\partial_z(\buvec{z}\buvec{\veckappa}_m+\buvec{\veckappa}_m\buvec{z}).
  \label{ch3:tensOmega}
  \end{equation}
  In the following, for a given vector $\bvec{V}$, the scalar quantities $V_s$, $V_{\kappa}$ and $V_z$ will refer to the components on $\buvec{s}_m$, $\buvec{\veckappa}_m$ and $\buvec{z}$ (respectively) of the vector $\bvec{V}$.
  
  Projecting eq. (\ref{Helmholtz reference appendix}) on $\buvec{s}_m$ gives 
  \begin{equation}
  (\partial_z^2 + (\gamma^o)^2) E^{ref}_{m,s}=0 \quad \text{with}  \quad (\gamma^o)^2=k_0^2 \epsilon^o-\kappa_m^2,   
  \label{Helmholtz TE}
   \end{equation}
  which corresponds to the transverse electric case. 
  
The projections of  eq. (\ref{Helmholtz reference appendix}) on $\buvec{\veckappa}_m$ and  $\buvec{z}$ couple $E^{ref}_{m,\kappa}$ and $E^{ref}_{m,z}$
\begin{eqnarray}
 \label{systErefb}  
  (-\partial_z^2-k_0^2\epsilon^o)E^{ref}_{m,\kappa}+i\kappa_m\partial_zE^{ref}_{m,z}=0 \\
  \label{systErefc}  
  (\kappa_m^2-k_0^2\epsilon^e)E^{ref}_{m,z}+i\kappa_m\partial_zE^{ref}_{m,\kappa}=0, 
\end{eqnarray}
Using the Maxwell equation $\nabla \times \Evec=i \omega \Bvec$, we can express the transverse magnetic field component $B^{ref}_{m,s}$ with respect to  $E^{ref}_{m,\kappa}$ and $E^{ref}_{m,z}$:
  \begin{equation}
  i\omega B^{ref}_{m,s}=i \kappa_m   E^{ref}_{m,z} - \partial_z E^{ref}_{m,\kappa}.
  \label{Bref1}
  \end{equation}
    Combining eq. (\ref{systErefb}), (\ref{systErefc}) and (\ref{Bref1}) we obtain
     \begin{equation}
    \quad (\partial_z \frac{1}{\epsilon^o} \partial_z+ \frac{(\gamma^e)^2}{\epsilon^e}) B^{ref}_{m,s}=0 \quad
 \text{with}  \quad (\gamma^e)^2=k_0^2 \epsilon^e-\kappa_m^2,
   \label{Helmholtz TM}
 \end{equation} 
  which corresponds to the transverse magnetic case. 
  
  %\begin{eqnarray}
  %\label{ch3:systErefa}  
  %\text{sur}\,\buvec{u}_0^1:& (\kappa_m^2-\partial_z^2-k_0^2\epsilon^o)E^{ref}_1=0 \\  
  %\label{ch3:systErefb}  
  %\text{sur}\,\buvec{u}_0^2:& (-\partial_z^2-k_0^2\epsilon^o)E^{ref}_2+i\kappa_m\partial_zE^{ref}_3=0 \\
%  \label{ch3:systErefc}  
   % \text{sur}\,\buvec{u}_0^3:& (\kappa_m^2-k_0^2\epsilon^e)E^{ref}_3+i\kappa_m\partial_zE^{ref}_2=0,
%  \end{eqnarray}

\subsection{The Green's tensor expressed as two scalar Green's functions}

The Green's tensor for the planar reference structure with a relative permittivity  anisotropic with symmetry axis $z$ can be expressed as  two scalar  Green's functions, also corresponding to the transverse electric and transverse magnetic cases. 
The Green's tensor is solution of 
\begin{equation}
  \opOmega_m(\tensG_m(z,z'))-k_0^2\tenseps^{ref}(z) \tensG_m(z,z')=k_0^2  \delta(z-z') \tensI. 
  \label{Helmholtz Green appendix}
  \end{equation}
  We note $G_m^{X,Y}$ the  $\buvec{X}\buvec{Y}$ component of the Green's tensor (with $\buvec{X}$ and $\buvec{Y}$ being equal to $\buvec{s}_m$, $\buvec{\veckappa}_m$ and $\buvec{z}$ successively).
  Using the expression of $\opOmega_m$ in the $(\buvec{s}_m,\buvec{\veckappa}_m,\buvec{z})$ basis leads to several results. 
  
First, it is possible to show that $G_m^{s,s}$ is solution of 
\begin{equation}
    (\partial_z^2 + (\gamma^o(z))^2)G_m^{s,s}=-k_0^2\delta(z-z'),
    \label{Helmholtz G11}
    \end{equation}
    which is the scalar equation for the transverse electric case.
    Second, we find that $G_m^{s,\kappa}$, $G_m^{s,z}$, $G_m^{\kappa,s}$ and    $G_m^{z,s}$ are equal to zero.
    Third, $G_m^{\kappa,\kappa}$ and  $G_m^{z,\kappa}$ are coupled by the following equations
        \begin{eqnarray}
    \label{eq22}
  (-\partial_z^2-k_0^2\epsilon^o(z))G_m^{\kappa,\kappa}+ i\kappa_m\partial_zG_m^{z,\kappa}=k_0^2\delta(z-z') \\
      \label{eq32}
  (\kappa_m^2-k_0^2\epsilon^e(z))G_m^{z,\kappa}+i\kappa_m\partial_zG_m^{\kappa,\kappa}=0, 
  \end{eqnarray}
    while    $G_m^{\kappa,z}$ and  $G_m^{z,z}$ are coupled by 
  \begin{eqnarray}
    \label{eq23}
  (-\partial_z^2-k_0^2\epsilon^o(z))G_m^{\kappa,z}+ i\kappa_m\partial_zG_m^{z,z}=0 \\
  \label{eq33}
  (\kappa_m^2-k_0^2\epsilon^e(z))G_m^{z,z}+i\kappa_m\partial_zG_m^{\kappa,z}=k_0^2\delta(z-z').
  \end{eqnarray}
Using eq. (\ref{eq22}) and (\ref{eq32}) and introducing 
\begin{equation}
G_m^{p,\kappa}=i\kappa_m G_m^{z,\kappa}-\partial_zG_m^{\kappa,\kappa},
\end{equation}
it is obtained that $G_m^{p,\kappa}$ is the solution of 
	\begin{equation}
	\left[\partial_z \frac{1}{\epsilon^o(z)} \partial_z+ \frac{(\gamma^e(z))^2}{\epsilon^e(z)} \right] G_m^{p,\kappa}=\partial_z\frac{k_0^2}{\epsilon^o(z)}\delta(z-z').
	\label{eq Gp2}
	\end{equation}
	From the eq. (\ref{eq22}) we express $G_m^{\kappa,\kappa}$  with respect to  $G_m^{p,\kappa}$
	\begin{equation}
	G_m^{\kappa,\kappa}=\frac{\partial_z G_m^{p,\kappa} -k_0^2\delta(z-z')}{k_0^2\epsilon^o(z)}.
	\label{eq Gm22}
	\end{equation}
	And from the eq. (\ref{eq32}), we express $G_m^{z,\kappa}$  with respect to  $G_m^{p,\kappa}$
	\begin{equation}
	G_m^{z,\kappa}=\frac{-i\kappa_m G_m^{p,\kappa}}{k_0^2\epsilon^e(z)}.
	\label{eq Gm32}
	\end{equation}
	
	Following the sames steps, using eq. (\ref{eq23}) and (\ref{eq33}) and introducing 
\begin{equation}
G_m^{p,z}= G_m^{z,z}-\frac{\partial_zG_m^{\kappa,z}}{i\kappa_m}
\end{equation}
it is obtained that $G_m^{p,z}$ is the solution of 
	\begin{equation}
	\left[\partial_z \frac{1}{\epsilon^o(z)} \partial_z+ \frac{(\gamma^e(z))^2}{\epsilon^e(z)}\right] G_m^{p,z}=-\frac{k_0^2}{\epsilon^e(z)}\delta(z-z').
	\label{eq Gp3}
	\end{equation}
	From the eq. (\ref{eq23}) we express $G_m^{\kappa,z}$  with respect to  $G_m^{p,z}$
	\begin{equation}
	G_m^{\kappa,z}=i\kappa_m\frac{\partial_z G_m^{p,z} }{k_0^2\epsilon^o(z)}.
	\label{eq Gm23}
	\end{equation}
	And from the eq. (\ref{eq33}), we express $G_m^{z,z}$  with respect to  $G_m^{p,z}$
	\begin{equation}
	G_m^{z,z}=\frac{\kappa_m^2 G_m^{p,z}-k_0^2\delta(z-z')}{k_0^2\epsilon^e(z)}.
	\label{eq Gm33}
	\end{equation}

The left member of the eqs. (\ref{eq Gp2}) and (\ref{eq Gp3}) is the operator involved in the equation for the transverse magnetic field problem (see eq. (\ref{Helmholtz TM})).
  Therefore, $G_m^{p,\kappa}$ and $G_m^{p,z}$ are two Green's functions, associated with the transverse magnetic field problem but with different sources (see the right member of eqs. (\ref{eq Gp2}) and (\ref{eq Gp3})).   In the following subsection, we deduce the link between  $G_m^{p,\kappa}$ and $G_m^{p,z}$ from the reciprocity principle. 
  
  \textbf{Note on the singularity of the Green's functions:}
  
  From  eq. (\ref{eq Gp2}), it appears that $\partial_z G_m^{p,\kappa}(z,z')$ presents a singularity equals to $k_0 ^2 \delta(z-z')$, from which we deduce, using eq. (\ref{eq Gm22}) that $G_m^{\kappa,\kappa}$ does not have any singularity at the interface $z=z'$. It also appears that $G_m^{p,\kappa}$ is non-singular, and from eq. (\ref{eq Gm32}) that $G_m^{z,\kappa}$ is also non-singular. From eq. (\ref{eq Gp3}), it appears that 
  $G_m^{p,z}$ and $\partial_z G_m^{p,z}(z,z')$ are non-singular. From eq. (\ref{eq Gm23}), we deduce that $G_m^{\kappa,z}$ is non-singular while from eq. (\ref{eq Gm33}), $G_m^{z,z}$ presents a singularity equals to $-\delta(z-z')/\epsilon^e(z)$. To sum up, we can write the Green's tensor $\tensG_m(z,z')$ separating the non-singular part $\tensG^{NS}_m(z,z')$ and the singularity: 
  \begin{equation}
\tensG_m(z,z')=\tensG^{NS}_m(z,z')-\frac{1}{\epsilon^e} \delta(z-z') \buvec{z}\buvec{z},
\label{G sing appendix}
\end{equation}

\subsection{Properties of the Green's functions related to the reciprocity principle}

We consider an Hilbert function space with an Euclidian scalar product defined by $\langle f|g \rangle=\int_{-\infty}^{\infty}\text{d}zf(z)g(z)$. The operators  $\mathcal{L}_s=\partial_z^2+(\gamma^o(z))^2$  and $\mathcal{L}_p=\partial_z \frac{1}{\epsilon^o(z)} \partial_z+ \frac{(\gamma^e(z))^2}{\epsilon^e(z)}$ in the left hand side of eqs. (\ref{Helmholtz G11}), (\ref{eq Gp2}) and (\ref{eq Gp3}) are self-adjoint.
Writing $\langle \mathcal{L}_s G_m^{s,s}|G_m^{s,s}\rangle =\langle G_m^{s,s}|\mathcal{L}_s G_m^{s,s}\rangle $ and using eq. (\ref{Helmholtz G11}), we deduce that
 \begin{equation}
G_m^{s,s}(z,z')=G_m^{s,s}(z',z).
\label{recip Gs}
\end{equation}
Writing $\langle \mathcal{L}_p G_m^{p,z}|G_m^{p,z}\rangle =\langle G_m^{p,z}|\mathcal{L}_p G_m^{p,z}\rangle $ and using eq. (\ref{eq Gp3}), we also deduce that 
\begin{equation}
\frac{G_m^{p,z}(z,z')}{\epsilon^e(z)}=\frac{G_m^{p,z}(z',z)}{\epsilon^e(z')}.
\label{recip Gp3}
\end{equation}
Last, writing $\langle \mathcal{L}_p G_m^{p,z}|G_m^{p,\kappa}\rangle =\langle G_m^{p,z}|\mathcal{L}_p G_m^{p,\kappa}\rangle $ and using eq. (\ref{eq Gp2}) and eq. (\ref{eq Gp3}), we deduce that 
\begin{equation}
G_m^{p,\kappa}(z,z')=\frac{\epsilon^e(z)}{\epsilon^o(z')} \partial_{z'} G_m^{p,z}(z',z),
\label{lien Gp3 Gp2}
\end{equation}
which, in combination with eq. (\ref{recip Gp3}), leads to
\begin{equation}
G_m^{p,\kappa}(z,z')=\frac{\epsilon^e(z')}{\epsilon^o(z')} \partial_{z'} G_m^{p,z}(z,z').
\label{lien Gp2 Gp3}
\end{equation}
These four relations are mathematical expressions for the consequences of the reciprocity principle on the Green's tensor, and they can be used to express $\tensG_m(z,z')$ with respect to $G_m^{s,s}$ and $G_m^{p,z}$ only.

\subsection{Expansion of the Green's tensor on its eigen modes}

\begin{itemize}
\item{\textit{ Transverse electric green function} }

The equation satisfied for the Green's function $G_m^{s,s}$ (transverse electric case) is
\begin{equation}
\left[\partial_z^2+ \left(\frac{2\pi}{\lambda}\right)^2 \epsilon^o(z)-\kappa_m^2 \right]G_m^{s,s}=-\left(\frac{2\pi}{\lambda}\right)^2\delta(z-z'),
\label{aux2}
\end{equation}
The homogeneous equation (eq. (\ref{Helmholtz TE})) for a mode with an electric field $\mathcal{E}^{n}(z)\buvec{s}_m$ and an eigen wavelength $\lambda_{n}$ can be written as
\begin{equation}
\frac{1}{\sqrt{\epsilon^o}}\left(\partial_z^2-\kappa_m^2\right)\frac{1}{\sqrt{\epsilon^o}}  (\sqrt{\epsilon^o}\mathcal{E}^{n})= - \left(\frac{2\pi}{\lambda_{n}}\right)^2 (\sqrt{\epsilon^o}\mathcal{E}^{n}). 
\label{eq mode s}
\end{equation}
For the sake of simplicity, we do not specify the dependence of the wavelength $\lambda_{n}$ and the field $\mathcal{E}^{n}(z)$ of the eigen mode on the subscript $m$ related to the in-plane wave vector $\veckappa_m$ considered in eq. \ref{aux2}. 

As $\frac{1}{\sqrt{\epsilon^o}}\left(\partial_z^2-\kappa_m^2\right)\frac{1}{\sqrt{\epsilon^o}}$ is a self-adjoint operator, its eigen modes form a basis and satisfy an orthogonality condition 
\begin{equation}
  \int_{z=-\infty}^{z=+\infty}\text{d}z \epsilon^o(z) \mathcal{E}^{n}(z)\mathcal{E}^{n'}(z) =\delta_{n,n'}.
 \label{eq norm s}
\end{equation}

We now want to expand $G_m^{s,s}$ on the basis formed by its eigen modes 
\begin{equation}
G_m^{s,s}(z,z')=\sum_n f_n(z') \mathcal{E}^{n}(z).
\end{equation}
Inserting this expression into eq. (\ref{aux2}), and using eq. (\ref{eq mode s}) we obtain
\begin{equation}
\sum_n f_n(z') \left[\left(\frac{2\pi}{\lambda}\right)^2-\left(\frac{2\pi}{\lambda_n}\right)^2\right] \epsilon^o(z)\mathcal{E}^{n}(z)=-\left(\frac{2\pi}{\lambda}\right)^2 \delta(z-z'), 
\end{equation}
from which we deduce, using the orthogonality condition (eq. (\ref{eq norm s}))
\begin{equation}
f_{n'}(z') \left[\left(\frac{2\pi}{\lambda}\right)^2-\left(\frac{2\pi}{\lambda_{n'}}\right)^2\right]  = - \left(\frac{2\pi}{\lambda}\right)^2 \mathcal{E}^{n'}(z').
\end{equation}
Hence, $G_m^{s,s}$ is expanded on the basis of its eigen modes
\begin{equation}
G_m^{s,s}(z,z')=\sum_n \frac{\mathcal{E}^{n}(z) \mathcal{E}^{n}(z')}{\left[\left(\frac{\lambda}{\lambda_n}\right)^2 -1 \right]}.
\label{exp G s}
\end{equation}

\item{\textit{Transverse magnetic Green's function}}

The  Green's function $G_m^{p,z}$ (transverse magnetic case) is the solution of 
\begin{equation}
\left[\partial_z \frac{1}{\epsilon^o}\partial_z+\frac{1}{\epsilon^e} \left(\left(\frac{2\pi}{\lambda}\right)^2\epsilon^e - \kappa_m^2\right)\right]G_m^{p,z}=-\frac{1}{\epsilon^e}\left(\frac{2\pi}{\lambda}\right)^2\delta(z-z').
\label{aux3}
\end{equation}
The homogeneous equation for a mode with a magnetic field $\mathcal{B}^{n}(z)\buvec{s}_m$ and an eigen wavelength $\lambda_{n}$ can be written as
\begin{equation}
\left[\partial_z \frac{1}{\epsilon^o}\partial_z-\frac{\kappa_m^2}{\epsilon^e} \right] \mathcal{B}^n= -\left(\frac{2\pi}{\lambda_n}\right)^2 \mathcal{B}^{n},
\label{eq mode p auto-adj}
\end{equation}
As $\left[\partial_z \frac{1}{\epsilon^o}\partial_z-\frac{\kappa_m^2}{\epsilon^e} \right]$ is a self-adjoint operator, its eigen modes form a basis and satisfy an orthogonality condition 
\begin{equation}
  \int_{z=-\infty}^{z=+\infty}\text{d}z (c\mathcal{B}^{n}(z))(c\mathcal{B}^{n'}(z)) =\delta_{n,n'},
 \label{eq norm p}
\end{equation}
where the speed of light $c$ in vacuum has been introduced so as to deal with quantities which have the unit of an electric field.
Following the same steps as for the $G_m^{s,s}$  function leads to
\begin{equation}
G_m^{p,z}(z,z')=\sum_n \frac{c^2\mathcal{B}^{n}(z)\mathcal{B}^{n}(z')}{\left[\left(\frac{\lambda}{\lambda_n}\right)^2 -1 \right] \epsilon^e(z')}.
\label{exp G p}
\end{equation}

\item{\textit{Green's tensor}}

We will use the results of Appendix C  to expand the Green's tensor on its eigen modes.
From eq. (\ref{lien Gp3 Gp2}), it comes
\begin{equation}
G_m^{p,\kappa}(z,z')= c^2\sum_n   \frac{\mathcal{B}^{n}(z)\partial_{z'} \mathcal{B}^{n}(z')}{\left[\left(\frac{\lambda}{\lambda_n}\right)^2 -1 \right] \epsilon^o(z')}.
\end{equation}
Then the equations relating $G_m^{z,z}$ and $G_m^{\kappa,z}$ to $G_m^{p,z}$ (eq. (\ref{eq Gm33}) and (\ref{eq Gm23})) give:  
\begin{equation}
	G_m^{z,z}(z,z')= c^2 \sum_n \frac{1}{\left[\left(\frac{\lambda}{\lambda_n}\right)^2 -1 \right]} \left[\frac{\kappa_m\mathcal{B}^{n}(z) }{k_0\epsilon^e(z)} \right] \left[\frac{\kappa_m\mathcal{B}^{n}(z') }{k_0\epsilon^e(z')} \right] -\frac{1}{\epsilon^e} \delta(z-z') \buvec{z}\buvec{z},
	\end{equation}
and
\begin{equation}
	G_m^{\kappa,z}(z,z')= c^2 \sum_n \frac{1}{\left[\left(\frac{\lambda}{\lambda_n}\right)^2 -1 \right]} \left[\frac{i\partial_z \mathcal{B}^{n}(z)}{k_0\epsilon^o(z)}  \right] \left[\frac{\kappa_m\mathcal{B}^{n}(z')}{k_0\epsilon^e(z')} \right].
	\end{equation}
Following the same steps, the equations relating $G_m^{z,\kappa}$ and $G_m^{\kappa,\kappa}$ to $G_m^{p,\kappa}$ (eq. (\ref{eq Gm32}) and (\ref{eq Gm22})) give:  
\begin{equation}
	G_m^{z,\kappa}(z,z')=-c^2 \sum_n \frac{1}{\left[\left(\frac{\lambda}{\lambda_n}\right)^2 -1 \right]} \left[\frac{\kappa_m\mathcal{B}^{n}(z) }{k_0\epsilon^e(z)} \right]\left[\frac{i\partial_{z'} \mathcal{B}^{n}(z')}{k_0\epsilon^o(z')}  \right],
	\end{equation}
and
\begin{equation}
	G_m^{\kappa,\kappa}(z,z')=-c^2\sum_n \frac{1}{\left[\left(\frac{\lambda}{\lambda_n}\right)^2 -1 \right]} \left[\frac{i\partial_z \mathcal{B}^{n}(z)}{k_0\epsilon^o(z)}  \right] \left[\frac{i\partial_{z'}\mathcal{B}^{n}(z')}{k_0\epsilon^o(z')}  \right].
	\end{equation}

\end{itemize}

To sum up, $\tensG_m^{NS}$ can be written in the form 
\begin{equation}
\tensG_m^{NS}(z,z')=\sum_n \frac{\Avec_{n}(z)\otimes\overline{\Avec_{n}}(z')}{\left[\left(\frac{\lambda}{\lambda_n}\right)^2 -1\right]},
\label{decomp tens G appendix}
\end{equation}
where $\otimes$ denotes the tensor product between two vectors, $\Avec_{n}(z)$ is defined by
\begin{equation}
\Avec_{n}(z)=\mathcal{E}_n(z) \buvec{s}_m -i c\left[ \frac{  -  \partial_z}{k_0\epsilon^o(z)} \buvec{\veckappa}_m+  \frac{i\kappa_m}{k_0\epsilon^e(z)}\buvec{z}\right]\mathcal{B}_n(z),
\end{equation}
while $\overline{\Avec_{n}}(z')$ is the complex conjugate of $\Avec_{n}(z')$.
From the Maxwell equation $\nabla \times \Hvec = \frac{\partial \Dvec}{\partial t}$, it is easy to show that $\Avec_{n}$ is the electric field of the mode.

\end{document}